\documentclass{aa}
  
\usepackage[varg]{txfonts}
\usepackage{graphicx}
\usepackage{xcolor}
\usepackage{multicol}
\usepackage{multirow}
\usepackage{array}
\usepackage{color}
\usepackage{caption}
\usepackage{ulem} 
\usepackage{subcaption}         
\usepackage{lscape}             
\usepackage{placeins}           

\begin{document}

\title{Characterization of the variability of the blue supergiant HD 14134}


 \author{Suryani Guha\inst{1,2} \corrauth{suryaniguha8@gmail.com}
 \and Michaela Kraus\inst{1}\email{michaela.kraus@asu.cas.cz}
 \and Julieta P. S\'{a}nchez Arias\inst{1} \email{julietapazsanchez@gmail.com}
 \and P\'{e}ter N\'{e}meth\inst{1} \email{pnemeth1981@gmail.com}
 \and Krzysztof Kami\'{n}ski\inst{3} \email{kkastr@gmail.com}
        }

   \institute{Astronomical Institute, Czech Academy of Sciences, Fri\v{c}ova 298, 251 65 Ond\v{r}ejov, Czech Republic
   \and Astronomical Institute, Faculty of Mathematics and Physics, Charles University, V Hole\v{s}ovick\'{a}ch 2, 182 00 Prague, Czech Republic
   \and Astronomical Observatory Institute, Faculty of Physics, A. Mickiewicz University, 
S\l{}oneczna 36, 60-286 Pozna\'{n}, Poland \\ }

   \date{Received xxx; accepted xxx}

 
  \abstract
   {The post-main sequence evolution of massive stars encompasses phases in which the stars display high variability. One such 
   class of objects are the blue supergiants. These objects may be in either the pre- or post-red supergiant phase of their 
   evolution. Their variability patterns might provide constraints for a proper classification of the objects.} 
   {The study aims to characterise the observed variability of the B-type supergiant HD~14134 and to
   investigate the imprint of a time-variable wind on the brightness variation of the star and its impact on the detectability of 
   pulsations signals.}
   {Spectroscopic data were collected over a five-month period and combined with space-photometry from the TESS mission.  
   Stellar parameters were derived from modelling of the time-averaged spectrum with CMFGEN and the spectral energy distribution 
   and were confirmed with stellar evolution models computed with MESA. The light curves and radial velocity curves of selected 
   lines were analysed with various methods to determine pulsation signals. The wind variability and its imprint on the stellar 
   brightness were investigated from an analysis of the H$\alpha$ line. Predictions of mode excitations were computed with the 
   GYRE pulsation code and compared to the frequencies determined from the observations.} 
   {A g-mode with a period of $\sim 19.2$\,d and its harmonics are consistently detected in all data sets.
    The spectra unveil strong, non-periodic wind variability and three frequency signals were identified as due to this wind 
    variability. The stellar parameters and age derived for HD~14134 together with the absence of radial pulsations classify 
    the star as a post-main sequence object evolving towards the red-supergiant stage, questioning its classification as 
    $\alpha$~Cyg variable.}
   {The results reinforce that simultaneous long-term spectroscopic and photometric monitoring is indispensable for reliable
   frequency detections and for disentangling of variabilities imprinted by a time-variable wind from those imposed by 
   pulsations.}

   \keywords{stars: massive -- supergiants -- techniques: spectroscopic -- techniques: photometric -- stars: individual: HD\,14134}

   \maketitle
   \nolinenumbers

\section{Introduction}

Blue supergiants (BSGs) are a class of massive stars whose evolutionary pathway is not well established. 
While they were supposed to be direct descendents of massive OB main-sequence stars, their relative numbers, 
especially for stars more massive than 30\,M$_{\odot}$ \citep{2014A&A...570L..13C}, seem to be in conflict 
with this interpretation as the only formation channel and challenges stellar evolution theories.

The observed overdensity in BSGs might be explained by two diverse populations predicted from stellar evolution 
models of single stars \citep[e.g.][]{2012A&A...537A.146E, 2021A&A...650A.128G}; (i) shell-hydrogen burning 
post-main sequence stars, and (ii) core-He burning post-red supergiants. On the other hand, there is growing 
evidence that the majority of massive stars are born in binary or multiple systems \citep{2012Sci...337..444S}, 
but the number of to date confirmed binaries among BSGs is small \citep{2023A&A...674A.212D, 
2024arXiv240511209S}. This might suggest that binary interaction may have taken place and that a significant 
number of luminous BSGs might be merger remnants \citep{2024ApJ...963L..42M, 2024A&A...690A..65H}. This 
scenario might also explain some of the surface abundance anomalies seen in BSGs \citep{2023Galax..11...93S, 
2024ApJ...963L..42M}. An apparent overpopulation of BSGs might also result from inaccuracies in stellar 
evolution models (core size, treatment of rotation, core-overshooting and mass-loss) that can lead to shifts in 
the position of the end of the main-sequence \citep[e.g.][]{2021A&A...648A.126M}. Each of these scenarios 
causes a different internal structure of the star. To distinguish BSGs resulting from the various formation 
channels, it is therefore necessary to gain insight into the stars' interiors.

High-precision space photometry 
and dedicated spectroscopic surveys
revealed that many BSGs pulsate \citep[e.g.][]{2006ApJ...650.1111S, 
2012A&A...542L..32K, 2017A&A...602A..32A, 2018MNRAS.476.1234A, 2018A&A...612A..40S, 2020A&A...640A..36B}. 
Theoretical investigations found an instability strip of gravity-mode pulsations in the BSG domain 
\citep{2006ApJ...650.1111S, 2013MNRAS.432.3153D} whose position and extent critically depends on the physical 
properties of the star during its main-sequence evolution \citep[core-overshooting, mass-loss, mixing, 
etc.][]{2009MNRAS.396.1833G, 2015MNRAS.447.2378O}. The type and amount of pulsation modes vary along 
the evolution and could be used to distinguish between BSGs in pre- and post-red supergiant phases 
\citep{2013MNRAS.433.1246S, 2019NatAs...3..760B}. Furthermore, differences in internal structure of BSGs resulting from single 
star and binary evolution (merger) channels may cause different pulsation patterns \citep{2024ApJ...967L..39B, 
2024A&A...690A..65H}. In addition to coherent pulsations, the amplitude spectra of many BSGs seem to contain 
stochastic low-frequency (SLF) variability \citep{2019NatAs...3..760B, 2020A&A...640A..36B, 2024ApJ...966..196M, 2025A&A...697A.152K}. The interpretation of this variability is yet unclear with suggestions ranging 
from internal gravity waves generated by core convection \citep{2019NatAs...3..760B}, to subsurface convection 
zones \citep{2021ApJ...915..112C}, surface turbulences \citep{2022ApJ...924L..11S}, and line-driven wind 
instabilities \citep{2021A&A...648A..79K}.

Many BSGs display time-variable winds \citep{1996A&A...305..887K, 2018A&A...614A..91H, 2023A&A...677A.176C, 2026A&A...710A.383C}, 
and a correlation between the occurrence of so-called strange modes
and phases of enhanced mass loss was proposed from observations \citep{2010A&A...513L..11A, 
2015A&A...581A..75K} and confirmed by theoretical modelling \citep[e.g. for the BSG star 
55\,Cygni,][]{2016MNRAS.457.4330Y}. Although strange-mode instabilities are predicted to occur in stars 
within the upper part of the Hertzsprung-Russell (HR) diagram spreading from hot to cool temperatures 
\citep{2024MNRAS.529.4947G}, not all excited modes automatically exceed the escape velocity and lead to 
mass-loss \citep{2021MNRAS.500.5515Y, 2026MNRAS.549ag895D}.

We aim to investigate the variability of the photosphere and winds of BSGs, as well as their possible 
connection to pulsations. For this, we focus on the Galactic BSG \object{HD 14134} (= 61~And; V520~Per, 
$V = 6.57$\,mag), which is classified as B3 supergiant with luminosity class Ia \citep{1955ApJ...122..429J, 
1992A&AS...94..569L, 2024arXiv240704163N}. The star is known for its variability in
(i) photometry \citep{1982A&AS...48..503R, 1990A&AS...83...11W, 1999A&A...345..505K, 2000IBVS.4946....1A, 
2002MNRAS.331...45K, 2009A&A...507.1141L, 2017A&A...598A.108L}, 
(ii) radial velocity measured in various photospheric lines \citep{1981A&AS...45..121K, 2024arXiv240511209S}, 
and 
(iii) its H$\alpha$ line, which displays a diversity of profile shapes such as P~Cygni, inverse P~Cygni, 
pure emission, pure absorption, and complete absence, i.e. compensation of absorption with emission 
\citep{2004MNRAS.351..552M, 2017RAA....17...38M, 2017JApA...38...20M}. 
It has been proposed to be a $\alpha$~Cyg variable with variability type IA 
\citep{2017A&A...598A.108L}, meaning it is an irregular early-type star.

We conducted spectroscopic observations of HD~14134 and supplemented 
the data with space-based photometry. The paper is organized as follows. 
In Sect.~\ref{sect:obs} we describe the spectroscopic and photometric data sets.
We determine the stellar parameters in Sect.~\ref{sect:stelparam} and analyse the line-profile, radial velocity, and wind variabilities detected in the spectroscopic time-series in Sect.~\ref{sect:variabilities}. The analysis of the photometric light curves is presented in Sect.~\ref{sect:photometry}. Our results are discussed in Sect.~\ref{sect:discussion} and the conclusions follow in Sect.~\ref{sect:conclusions}.

\begin{table*}[ht!]
\centering    
\caption{Stellar and wind parameters of HD~14134.}   
\label{tab:stelparam}
\begin{tabular}{c|cccc}      
\hline \hline 
Parameter           &   \protect{\citet{1999A&A...349..553M}}   &   \protect{\citet{2006A&A...446..279C}}  &  \protect{\citet{2024A&A...687A.228D}}  & This Work\\
                    &      TLUSTY &   CMFGEN  & FASTWIND & CMFGEN \\
\hline                       
$T_{\rm eff}$ [K]      & 18000$\pm$1000  &  16000$\pm$1000 & 16700$^{+400}_{-300}$ & 16100$^{+260}_{-150}$ \\
$\log{g}$ [cgs]        & 2.3$\pm$0.2     &   2.05           & 2.17$\pm$0.05 &  2.09$\pm$0.04\\
$R$  [R$_{\odot}$]           & 48.5 &  56.7 & -- &  48.6$\pm$1.0 \\
$\log L/\textrm{L}_{\odot}$  & 5.35 &  5.28 & -- &  5.16$\pm$0.03 \\
$M$ [M$_{\odot}$]       & --  &  13.2  & -- & 10.6$\pm$1.0 \\
$v\sin{i}$  [km/s]     & 60$\pm$20  &  66  & 37 &  37$\pm$5 \\
$v_{\rm mac}$ [km/s]   & --  & --  &  51 &  51$\pm$4 \\
$v_{\rm mic}$ [km/s]   & 10$\pm$5  &  15 & 14$\pm$1 &  14$\pm$3 \\
$\beta$                & --  &  2   & 1  &  2-3 \\
$v_\infty$  [km/s]     & --  &  465 & -- &  590$\pm$40 \\
$\log{\dot{M}}$ [M$_\odot$/yr]   & -- &   -6.28 & -- &  -6.70$\pm$0.15 \\
$\log{Q}$                        & -- &  -12.9    & -13.3$\pm$0.1  & -13.38$\pm$0.1\\
\hline
 \end{tabular}
 \tablefoot{The parameters to compute $Q$ are $\dot{M}$ (M$_{\odot}$/yr), $v_\infty$ (km\,s$^{-1}$), and $R$ (R$_{\odot}$).}
\end{table*}

\section{Observations and Data Reduction}\label{sect:obs}

\subsection{Spectroscopy}

Spectroscopic monitoring was carried out from 2017 October 11 to 2018 March 12 utilizing the 
observing facilities of Poznan Spectroscopic Telescope 2 \citep[PST2,][]{2014IAUS..301..437K}, which is situated at the
Winer Observatory, Arizona, USA. A total of 146 high-resolution ($R \approx 40\,000$) spectra were 
obtained with a wavelength coverage of 3877 - 7114\,\AA. During three nights, the set-up was 
slightly shifted to longer wavelengths, covering a range of 4325 - 7725\,\AA. The exposure time 
per spectrum was in most cases 1200\,s, in four nights with worse observing conditions it was 
increased to 1800\,s. The spectra were reduced, heliocentric velocity corrected, and normalized to 
the continuum using standard \textsc{iraf}\footnote{IRAF was written at the National  Optical Astronomy 
Observatory, which was operated by the Association of Universities for Research in Astronomy (AURA) under 
cooperative agreement with the National Science Foundation.} \citep{1986SPIE..627..733T, 1993ASPC...52..173T} 
tasks. Whenever possible, two consecutive spectra were taken each night. These were combined to correct 
for cosmic rays and to increase the signal-to-noise ratio (S/N). The final 73 
individual spectra (see Table~\ref{table:obs} for details) have S/N values ranging from 30 to 110.

\subsection{TESS photometry}

The Transiting Exoplanet Survey Satellite (TESS) mission of NASA \citep{2014SPIE.9143E..20R} 
facilitates the high-precision photometric survey of the sky. It provides the time-series data 
products to study the variability of celestial objects. HD~14134 (TIC 264730431) was 
observed twice, in sector 18 (2--27 November 2019), and again, almost three years later, in sector 58 (29 
October - 26 November 2022). The time-series were sampled with a cadence of 1800\,s and 200\,s, respectively. 
Both data sets were taken with camera 2 and CCD 4. We note that HD~14134 was also observed in sector 85, 
however, the data seem to be corrupted and do not allow to construct a reliable light curve for our target. 
Hence we have omitted them from our analysis.

We downloaded $20 \times 20$\, pixel cutouts from the Full Frame 
Images (FFIs) and utilized the python package \textsc{lightkurve} 
\citep{2018ascl.soft12013L} to extract the light curves from them. 
For these FFI cutouts,
there are no predefined apertures available from the Science Processing 
Operations Center \citep[SPOC,][]{2016SPIE.9913E..3EJ}. To avoid contamination of the light curves 
with close-by background (or foreground) stars \citep{2023AJ....165..141H, 2023AJ....165..239P}, we consulted the
Gaia database \citep{2018A&A...616A...2L} for precise positions and magnitudes of stars within our 
field-of-view (FOV). We have used Gaia Data Release (DR) 2 instead of the more recent DR3,
as the latter one provides considerably less stars in our FOV. 

The apertures were selected to maximize the flux while minimizing contamination from other stars in the field. 
For the latter, we tested that background stars in the selected pixels do not add variable signals to the light 
curve of HD~14134. Background pixels are chosen to gauge background contamination (see 
Fig.\,\ref{fig:TPF_sectors} for details). After subtraction of the background flux, the stellar flux is 
converted to magnitudes and normalized to the mean value of the entire light curve within one sector. The 
final light curves for both sectors are shown in Fig.~\ref{fig:lc}.

\subsection{Gaia photometry}

We have obtained multi-band photometric data from Gaia DR3 \citep{2022gdr3.reptE....V}. To date, 
49 measurements have been released for HD~14134 (DR3 458374788336739200) spreading from 2014 
August 30 to 2017 March 20. The observations were carried out in the bands $G$ (330-1050 nm), 
$G_{\rm BP}$ (330-680 nm) and $G_{\rm RP}$ (680-1050 nm). Figure~\ref{fig:gaia} displays the Gaia 
light curves in magnitudes in the different bands (top panel) and the temporal colour variation 
($G_{\rm RP} - G_{\rm BP}$, bottom panel).


\section{Stellar parameters}\label{sect:stelparam}

\begin{figure*}[ht!]
  \centering
  \includegraphics[width=\textwidth]{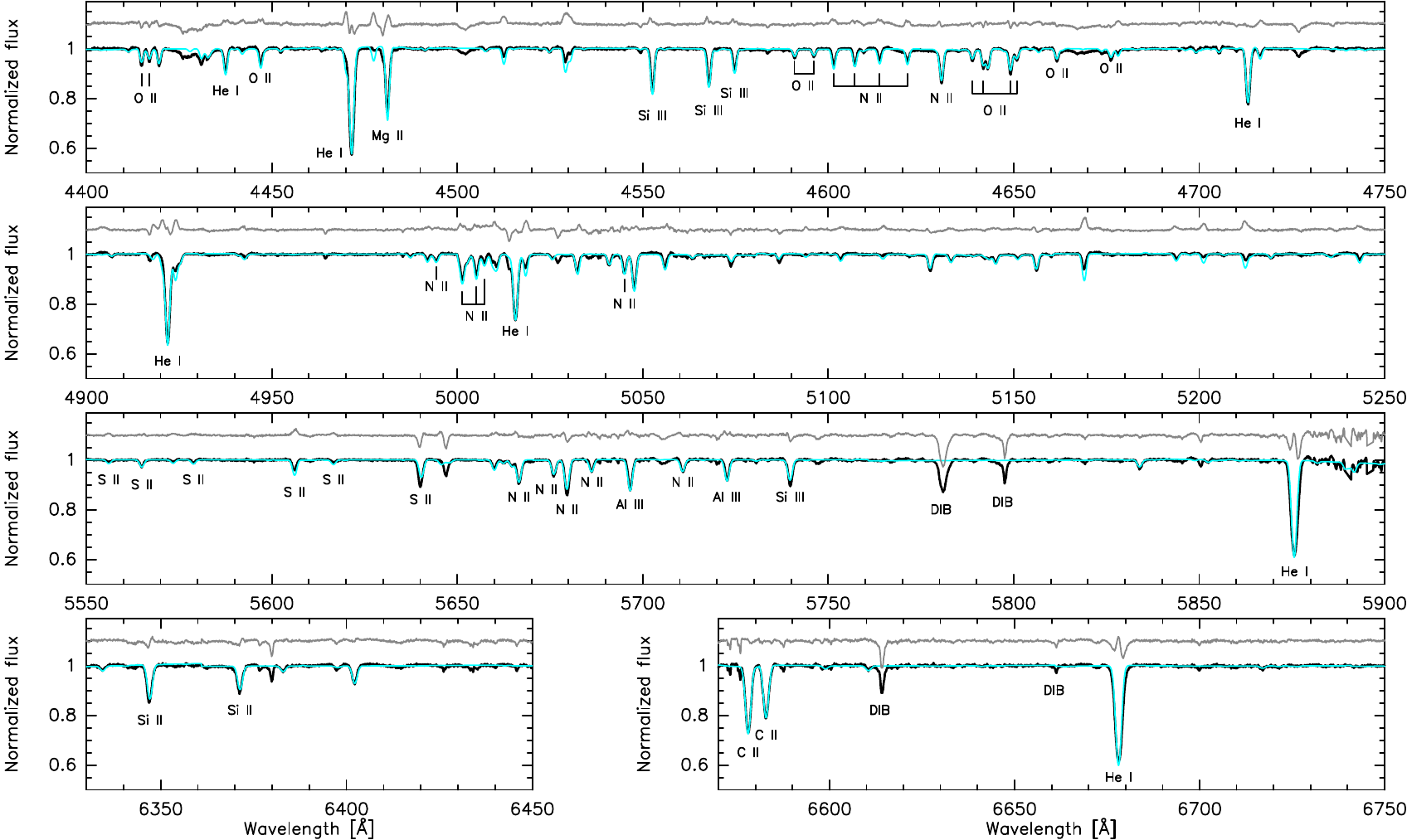}
  \caption{Fit (blue line) to the averaged spectrum (black line) of HD\,14134. The CMFGEN model  
  was calculated with the photospheric and wind parameters listed in Table~\ref{tab:stelparam}.
  For better visibility, the residuals (in gray) are offset by 0.1 in flux in each panel.}
  \label{fig:xtgrid01} 
\end{figure*}

Stellar and wind parameters of HD~14134 were determined in the literature based on non-LTE modelling using 
different codes, and Table~\ref{tab:stelparam} provides an overview of the most relevant studies. The values 
for the stellar parameters agree fairly well except for the stellar rotation projected to the line of sight, 
($v \sin i$). While older studies only used $v \sin i$ to broaden the photospheric lines, a combination of 
$v \sin i$ and the so-called macroturbulent velocity ($v_{\rm mac}$) seems nowadays to be the better approach 
\citep{2007A&A...468.1063S, 2014A&A...562A.135S}. The cause of macroturbulence within the atmospheric layers 
of supergiants is still not fully understood but was proposed to be related to pulsation activity 
\citep{2009CoAst.158...66A, 2010AN....331.1069S}.
 
Our spectra of HD\,14134 (see Fig.\,\ref{fig:xtgrid01}) display photospheric absorption lines in agreement with its classification as an early 
B-type supergiant \citep[e.g.][]{1990PASP..102..379W}. In addition, we observe emission components of the hydrogen Balmer lines, in particular in 
the lower Balmer series. These show rapid and substantial changes likely due to wind variability, which is 
further discussed in Sect.\,\ref{sect:wind}. We also observe line-profile variability in the photospheric 
lines, which we discuss in Sect.\,\ref{sect:var_stelparam}. Given these variabilities, we aim to reproduce the 
averaged spectrum of HD~14134 to derive average stellar parameters. 

We utilized synthetic model atmosphere spectra calculated with the non-LTE radiative transfer code CMFGEN 
\citep{1998ApJ...496..407H}. To limit the computational cost, several atmospheric parameters were fixed during 
the main analysis. The abundances of He, Ne, Mg, Al, Si, S, Fe, and Ni were initially set to solar values, and 
the microturbulent velocity ($v_{\rm mic}$), $v_{\rm mac}$, and $v \sin i$ were adopted from 
\cite{2024A&A...687A.228D}, see Table~\ref{tab:stelparam}. The free parameters include the effective temperature ($T_{\rm eff}$), surface 
gravity ($\log g$), CNO abundances, stellar radius ($R$) and mass-loss rate ($\dot{M}$). The wind 
structure follows a $\beta$-type velocity law with the terminal wind velocity ($v_\infty$) and $\beta$ as 
additional free parameters. 

\begin{figure}[t!]
  \centering
  \includegraphics[width=0.9\columnwidth]{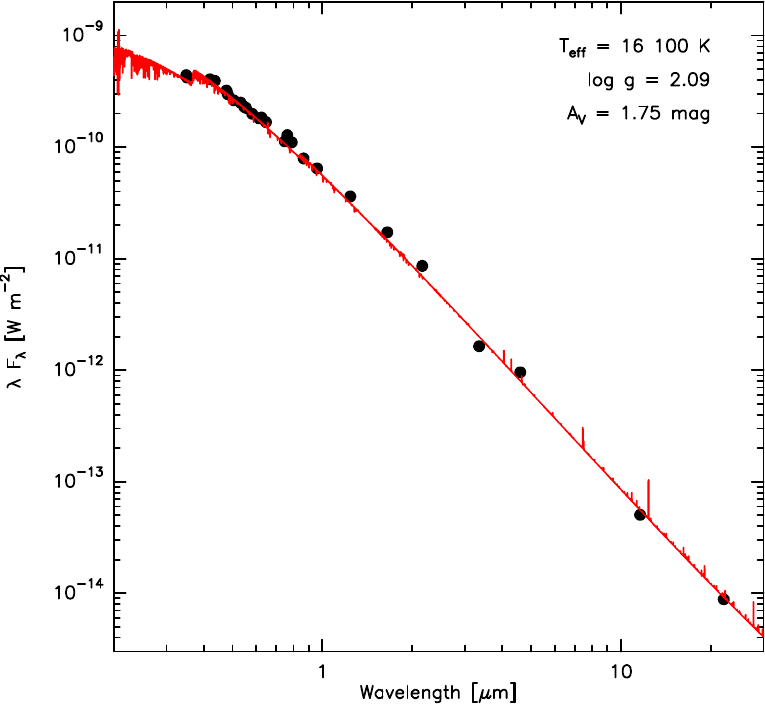}
  \caption{Fit of the CMFGEN model to the SED of HD~14134.}
  \label{fig:sed} 
\end{figure}

We initiated the spectral analysis using the parameter set of \cite{2024A&A...687A.228D} as listed in 
Table~\ref{tab:stelparam}, adopting homogeneous winds throughout. The primary goal was to reproduce the observed 
photospheric variability, which requires a consistent treatment of the stellar wind. In contrast to the FASTWIND models computed by \cite{2024A&A...687A.228D}, which parametrize the wind via the optical depth invariant $Q=\dot{M}\,(v_\infty\cdot R)^{-3/2}$, CMFGEN 
treats $\dot{M}$, $v_\infty$ and $R$ as independent quantities. We started with the literature value for the 
terminal wind velocity of $v_\infty = 465$\,km\,s$^{-1}$ obtained by \citet{1997MNRAS.284..265H}. We 
derived the stellar radius from simultaneous fitting of the spectral energy distribution (SED) of HD~14134 with 
the spectrum computed with CMFGEN. The SED in the UV and optical spectral range is only sensitive to 
$T_{\rm eff}$ and $\log g$, so that the extinction can be determined. Using the most recent distance estimate 
of $2.11\pm 0.21$\,kpc \citep{2025MNRAS.tmp.1439G} delivers the stellar radius, 
luminosity, and the mass of the star. These values serve as new input to CMFGEN. The entire process is iterated 
until both the spectrum and the SED are matched with a homogeneous set of parameters.

The spectral modelling was carried out using a steepest-descent $\chi^{2}$ minimization procedure implemented in 
{\sc XTgrid} \citep{2012MNRAS.427.2180N, 2019ASPC..519..117N}. New models were computed iteratively along the 
descent path, avoiding interpolation within precomputed grids and allowing for the simultaneous optimization of 
multiple parameters. The resulting spectra are broadened with the values of $v_{\rm mac}$ and $v \sin i$ and 
convolved to the resolution of the observed spectra. We explored the degeneracy between macroturbulent and 
rotational broadening. A total broadening of $63$\,km\,s$^{-1}$ reproduces the observed line profiles in the 
averaged spectrum, with the best agreement obtained for $v \sin i = 37$\,km\,s$^{-1}$ and $v_{\rm mac} = 
51$\,km\,s$^{-1}$ in agreement with the values found by \cite{2024A&A...687A.228D}. Finally, we refined the 
abundances of Mg, S, and Fe, finding that slightly subsolar values ($-0.1$\,dex) improve the quality of the 
fit. Slightly subsolar abundances were also found for the star's host cluster h~Per 
\citep{2021MNRAS.504..356D} supporting our findings.

The parameters of our best-fitting model are included in Table~\ref{tab:stelparam}. Inspection of 
Fig.~\ref{fig:xtgrid01} shows decent agreement between the synthetic (blue) and the observed (black) 
photospheric metal lines in the averaged spectrum. The goodness of the fit can also be assessed from the 
residuals (plotted in gray), which are shifted upward in flux for better visibility.
The fit to the SED is shown in Fig.\,\ref{fig:sed}. The observations were taken from the HIPPARCOS and 
TYCHO catalogues\footnote{\url{https://cdsarc.cds.unistra.fr/viz-bin/cat/I/239}} \citep{1997ESASP1200.....E}, 
Gaia DR3\footnote{\url{https://cdsarc.cds.unistra.fr/viz-bin/cat/I/355}} \citep{2023A&A...674A...1G}, the 
WISE All-Sky Data Release\footnote{\url{https://cdsarc.cds.unistra.fr/viz-bin/cat/II/311}} 
\citep{2012wise.rept....1C}, and the Pan-STARRS1 
release\footnote{\url{https://cdsarc.cds.unistra.fr/viz-bin/cat/II/349}} \citep{2016arXiv161205560C}.
We obtain an interstellar extinction value of $A_{V} = 1.75\pm 0.02$, in good agreement with the value 
reported by \citet{2021MNRAS.504..356D} for the cluster h~Per in which HD~14134 resides.

From the spectral fitting, we find slightly lower 
values for the effective temperature and surface gravity compared to \citet{2024A&A...687A.228D} but more
closely matching the results of \citet{2006A&A...446..279C}. \citet{2006A&A...446..279C} derived 
$\textrm{[N/C]} = +0.8$ and $\textrm{[N/O]} = +0.9$ based on \citet{2006NuPhA.777....1A}, while we obtain 
slightly lower values of $\textrm{[N/C]} = +0.6 \pm 0.1$ and $\textrm{[N/O]} = +0.8 \pm 0.1$ using the solar 
abundances of \citet{2009ARA&A..47..481A}. We observe an extra broadening of the He lines. 
This suggests that the He lines might require higher values of $v_{\rm mac}$ as observed in other B-type supergiants 
\citep{2015A&A...581A..75K, 2018A&A...614A..91H} or that $v_{\rm mac}$ is time-variable as found for 55~Cygnus 
\citep{2023A&A...677A.176C} or for the rapidly rotating massive star $\zeta$~Oph \citep{2025A&A...703A...2K}.

\begin{figure}[t!]
  \centering
  \includegraphics[width=\columnwidth]{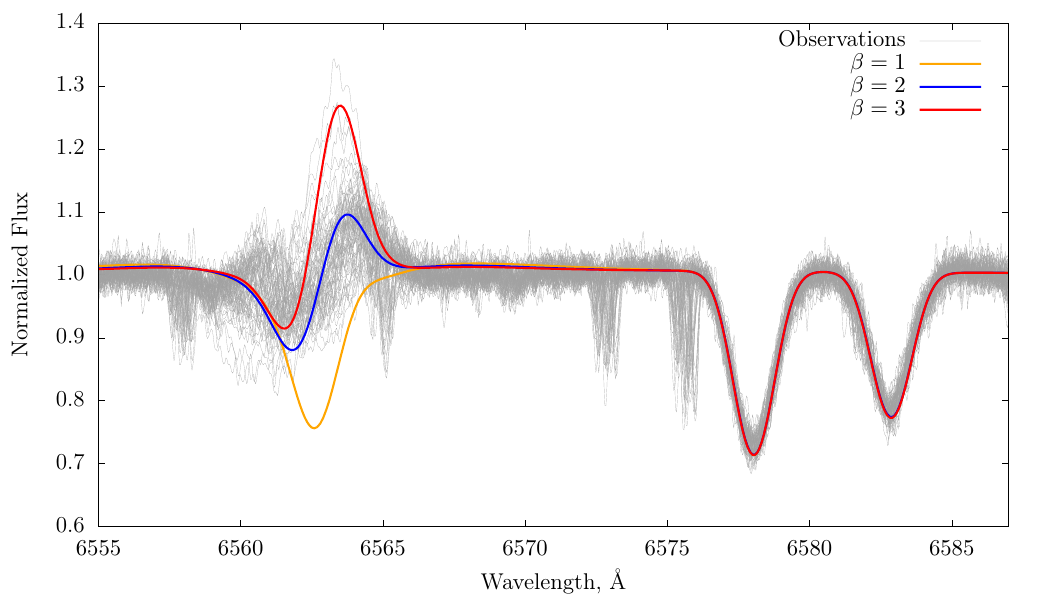}
  \caption{Comparison of models with varying $\beta$-parameter to the entire sample of observed H$\alpha$ profiles.}
  \label{fig:beta} 
\end{figure}

Furthermore, it became evident early in the analysis that $\beta = 1$ does not reproduce the observed line 
profiles of the Balmer lines. This may be due to the fact that the average spectrum includes all (non-periodic) 
wind variations and, therefore, is inadequate to represent the stellar spectrum at any single (snapshot) time. 
Our model with $\beta = 1$ results in an H$\alpha$ line in absorption\footnote{We note that 
\citet{2024A&A...687A.228D} provide neither information on the profile shape of H$\alpha$ of the spectrum they 
have modelled nor whether they could achieve a decent fit to it with their best-fitting parameter set.} while 
most of our spectra show H$\alpha$ in emission, often with a P~Cygni, sometimes a double-peaked profile (see 
Fig.\,\ref{fig:Ha_profile}). Clearly, a higher value of $\beta$ is needed to reproduce at least some of the 
individual observations (Fig.\,\ref{fig:beta}). For most epochs, the Balmer lines favour $\beta \simeq 2$ in 
agreement with earlier findings by \citet{2006A&A...446..279C}, while episodic phases require significantly 
higher values ($\beta > 3$), indicating variability in the wind acceleration. However, with such a highly 
dynamical wind as we observe for HD~14134, the approach of modelling any individual snapshot observation with a 
stationary atmosphere plus wind model, as computed with CMFGEN or comparable codes, to derive mass-loss rates 
and wind parameters seems to be at least questionable.

\begin{figure}
    \centering
    \includegraphics[width=\columnwidth]{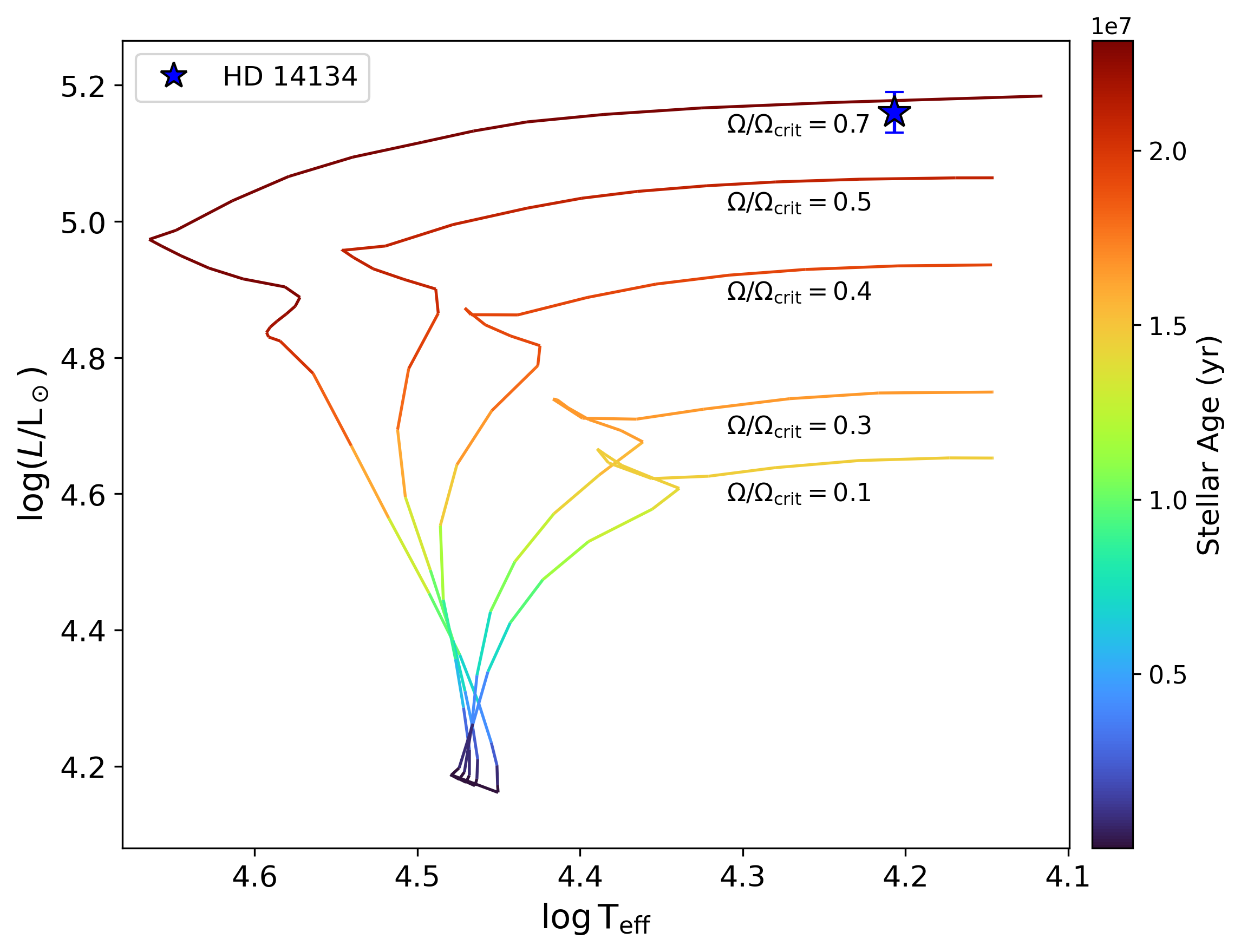}     
    \caption{MESA evolutionary models for a $14$\,M$_{\odot}$ star with different initial rotation velocities.}
    \label{fig:HR-mass}
\end{figure}

\subsection{Stellar evolution model for HD~14134}

The stellar parameters obtained from SED modelling and spectral fitting (temperature, luminosity, and mass, see Table\,\ref{tab:stelparam}), as well as the age of the star that is constraint by the age of the cluster, for which values in the range $10-30$\,Myr are reported \citep{2021MNRAS.504..356D,
2026RAA....26d5016T}, are compared with the predictions from stellar evolution models.
For this, we calculated evolutionary tracks with the Modules for Experiments in Stellar Astrophysics 
\citep[MESA r23.05.1,][]{2023ApJS..265...15J}
for solar metallicity stars ($Z=0.0142$). 

We first focused on a pre-red supergiant scenario and developed the models from the zero-age main sequence beyond core-hydrogen burning and until the temperature of HD~14134 is reached. The set-up of our models and values used for specific parameters, including numerical convergence constraints and time-steps, is in line with other works on massive stars \citep[see e.g.][]{2013ApJS..208....4P, 2025ApJ...994...77H}.
More specifically, we used the Ledoux criterion for the treatment of convective boundaries. Convective mixing is addressed using the time-dependent convection theory  with a mixing length parameter of $\alpha_{\rm MLT} = 1.5\,H_{p}$, where $H_{p}$ is the pressure scale height at the outer boundary of the core. 
We employed step-overshooting for the hydrogen-burning convective core with an overshooting parameter of $\alpha_{\rm ov} = 0.18\,H_{p}$, similar to the value adopted by \citet{2021A&A...648A.126M}.
Moreover, semi-convection was included with an efficiency of $\alpha_{\rm sc} = 0.01$  
allowing for moderate chemical mixing in Schwarzschild-unstable but Ledoux-stable zones. 
Furthermore, we used the updated and expanded OPAL equation-of-state tables from \citet{2002ApJ...576.1064R} and nuclear reaction rates from the Nuclear Astrophysics Compilation of Reaction rates (NACRE; \citealt{1999NuPhA.656....3A}) and the Joint Institute for Nuclear Astrophysics Reaction Library (JINA REACLIB database; \citealt{2010ApJS..189..240C}).
We included the solar heavy elements mixture given by \citet{2009ARA&A..47..481A}, and the OPAL opacity tables
\citep{1996ApJ...464..943I} for the higher-temperature regime.
For mass loss along the evolution, the prescriptions by \citet{2001A&A...369..574V} and \citet{2002A&A...389..162N} are used with a scaling factor of $1.0$. They are valid for temperatures $T_{\rm eff} > 10\,000$\,K and for surface hydrogen abundances of $X(H) > 0.4$ and $X(H) < 0.4$, respectively.
\begin{figure}
    \centering
    \includegraphics[width=\columnwidth]{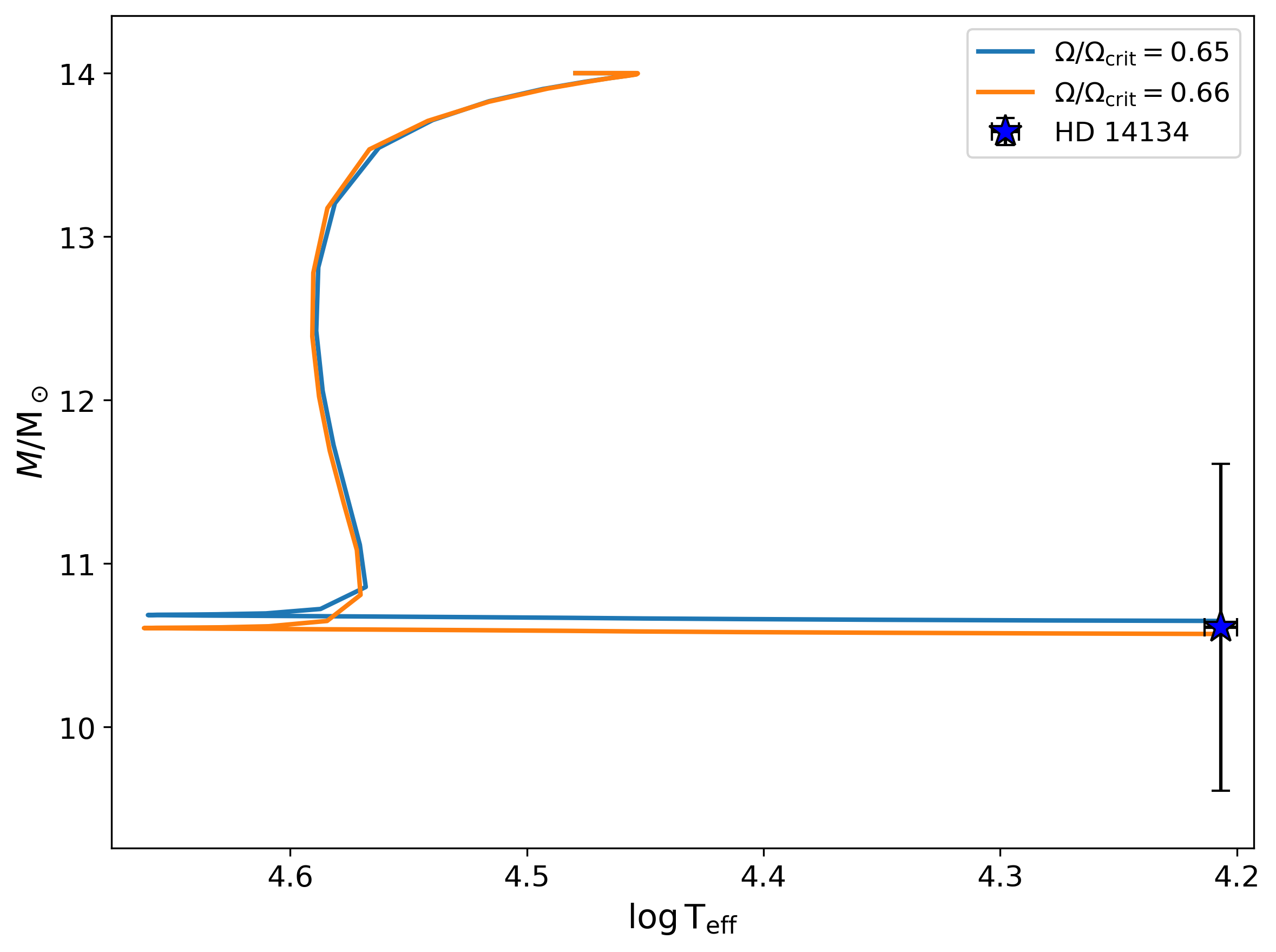}
    \caption{Evolution of the stellar mass for models with an initial mass of $14$\,M$_{\odot}$ and two different rotation velocities that closest reproduce the observed parameters of HD~14134.}
    \label{fig:mass}
\end{figure}

\begin{table}[t!]
\centering
\small
\caption{Parameters of stellar models derived from MESA for a $14\,$M$_\odot$ star with different initial rotation rates.}
\begin{tabular}{lcc}
\hline
\hline
Parameter & M1  & M2  \\
 & ($\Omega/\Omega_{\rm crit}=0.65$) & ($\Omega/\Omega_{\rm crit}=0.66$) \\
\hline
Initial Mass (M$_\odot$) & 14 & 14 \\
Current Mass (M$_\odot$) & 10.65 & 10.57 \\
Age (Myr) & 22.99 & 23.04 \\
T$_{\rm eff}$ (K) & 16087.0 & 16099.6\\
Radius (R$_\odot$) & 50.18 & 49.52 \\
Luminosity ($\log L/\textrm{L}_{\odot}$) & 5.18 & 5.17 \\
\hline
\end{tabular}
\label{tab:mesa_models}
\end{table}

As mentioned by \citet{2000ARA&A..38..143M}, high initial rotation causes the star to evolve to higher luminosities. This means that fast rotators are over-luminous with respect to their actual masses. 
Accordingly, the high luminosity observed in HD 14134 compared to its relatively low mass could suggest that the star originally rotated rapidly.
Therefore, we included rotation into our models. The chemical mixing and angular momentum transport due to the rotationally induced instabilities are treated in a diffusion approximation \citep{2013ApJS..208....4P}. For the efficiency of the total rotation induced composition mixing we use $f_{c} = 1/30$ 
and for the sensitivity of the rotationally induced mixing to the mean
molecular weight gradient we use $f_{\mu} = 0.05$  following \citet{2000ApJ...528..368H}.

With this defined set-up, we computed evolutionary tracks for stars in the mass range $11 - 15$\,M$_{\odot}$ and for a large range of initial rotation velocities. A general trend observed is that with higher initial rotation given in terms of the critical rotation ($\Omega/\Omega_{\rm crit}$), the star evolves to higher luminosities along the main sequence. Moreover, if the initial rotation velocity reaches (and exceeds) a value of $\Omega/\Omega_{\rm crit} \sim 0.5$, the star starts to develop a chemically homogeneous evolution due to the strong mixing. As a consequence, it becomes hotter during the main sequence as can be seen in the top panel of Fig.\,\ref{fig:HR-mass} for the example of models with increasing initial rotation and an initial mass of $14$\,M$_{\odot}$. 
While the effect of chemically homogeneous evolution is considerably more pronounced at lower metallicities \citep{2011A&A...530A.115B, 2015A&A...581A..15S}, it has also been reported to become dominant for rapidly rotating massive stars with solar metallicity using the Bonn models \citep{2013A&A...554A..23M} and MESA \citep{2025ApJ...994...77H}, so that we can be confident in the validity of our results.

\begin{figure*}
    \centering
    \includegraphics[width=\textwidth]{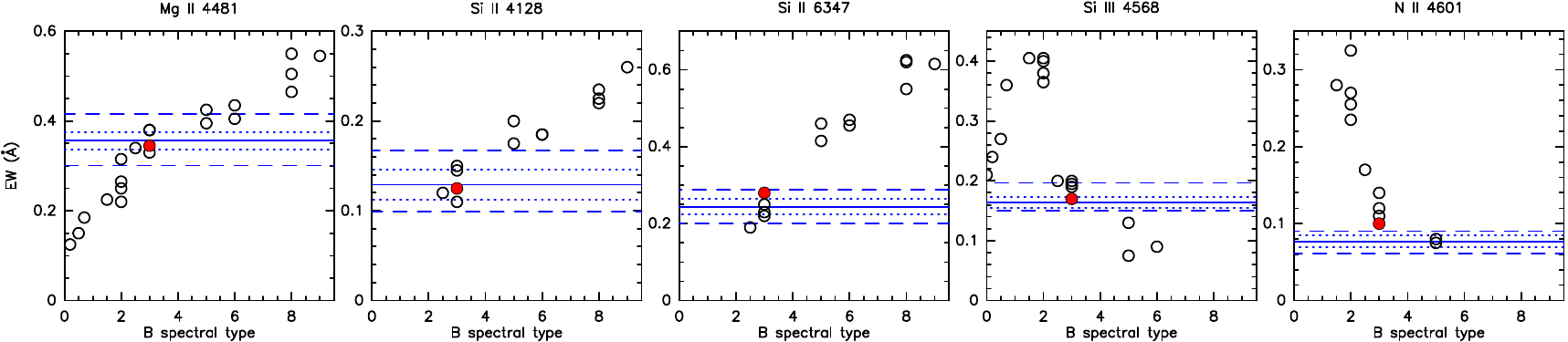}
    \caption{EWs of temperature sensitive photospheric lines in Galactic B-type supergiants (limited to luminosity class Ia) from \citeauthor{1993A&AS...97..559L} (\citeyear{1993A&AS...97..559L}, open black circles) along with our measurements for HD~14134 (filled red circle). Shown are the mean values (solid blue lines) and their standard deviations (dotted blue lines) as well as the observed range limited by the minimum and maximum values (dashed blue lines). The red filled circles are the measurements for HD~14134 from \citet{1993A&AS...97..559L}.}
    \label{fig:EW}
\end{figure*}

The stellar evolution models that most closely match all observed parameters of HD~14134 correspond to a star with an initial mass of $\sim 14$\,M$_{\odot}$ and a rotation rate of $\Omega/\Omega_{\rm crit}=0.675-0.680$. 
The mass evolution of the two boundary models is shown in the lower panel of Fig.\,\ref{fig:mass}. And the extracted stellar parameters for them are listed in Table\,\ref{tab:mesa_models}. These evolutionary models are further explored in Sect.\,\ref{sect:puls_wind}, where they serve as input for the computations of pulsations.
We also explored slightly higher or lower initial mass models, but they matched either the luminosity of the star or its mass, but not both simultaneously. Therefore, we consider the models presented in Table\,\ref{tab:mesa_models} as representative for HD~14134.

We also explored whether HD~14134 could be in the post-red supergiant evolution. To test this possibility, we developed MESA models for stars with higher initial masses and for a range of rotation velocities. We started from an initial mass of $20$\,M$_{\odot}$, which serves as sort of a lower limit for the post-red supergiant evolution, because stars with initial masses $8 - 20$\,M$_{\odot}$ end their lives as red supergiants and do not return \citep{2009ARA&A..47...63S, 2012A&A...537A.146E}. However, we discarded the post-red supergiant scenario for HD~14134, because all models evolved to significantly higher luminosities compared to the observed value.

\section{Line-profile and radial velocity variabilities}\label{sect:variabilities}

\subsection{Equivalent width variations}
\label{sect:var_stelparam}

Photospheric lines in the stellar spectrum carry information about the star's effective temperature 
and surface gravity. \citet{1993A&AS...97..559L} measured the equivalent widths (EWs) of 
hydrogen, helium, and metal lines (C, N, O, Mg, Si, Ne, Fe, in different ionization states) in the optical 
spectral range of late O to early A-type giants and supergiants and showed that they may change as functions of 
spectral type. In a later work, \citet{2007A&A...463.1093L} measured EWs of three selected photospheric
lines (He\,{\sc i} 4471\,\AA, Si\,{\sc ii} 4128\,\AA, Si\,{\sc iii} 4552\,\AA) in synthetic spectra of B-type 
supergiants computed with FASTWIND and presented them in isocontour plots as function of $T_{\rm eff}$ and 
$\log g$. Both studies show that the equivalent widths can be very sensitive even to small changes in effective 
temperature.

We measured the EWs of the photospheric lines Si\,{\sc ii}\,$\lambda\lambda 4128,6347$,
Si\,{\sc iii}\,$\lambda 4568$, N\,{\sc ii}\,$\lambda 4601$, and Mg\,{\sc ii}\,$\lambda 4481$, which are 
sensitive to the effective temperature of early B-type supergiants. Individual measurements are listed in 
Table \ref{tab:Eq_width}, and the mean values and their standard deviation are given in the bottom rows of 
Table \ref{tab:Eq_width}. To minimize the errors, we only considered spectra with local values of S/N $\ge$ 50 
for the EW measurements. Errors are typically on the order of $5-10\%$ for spectra with S/N higher than $\sim 
70$, and about $15\%$ for the lower quality ones.

\begin{figure}
    \centering
    \includegraphics[width=\columnwidth]{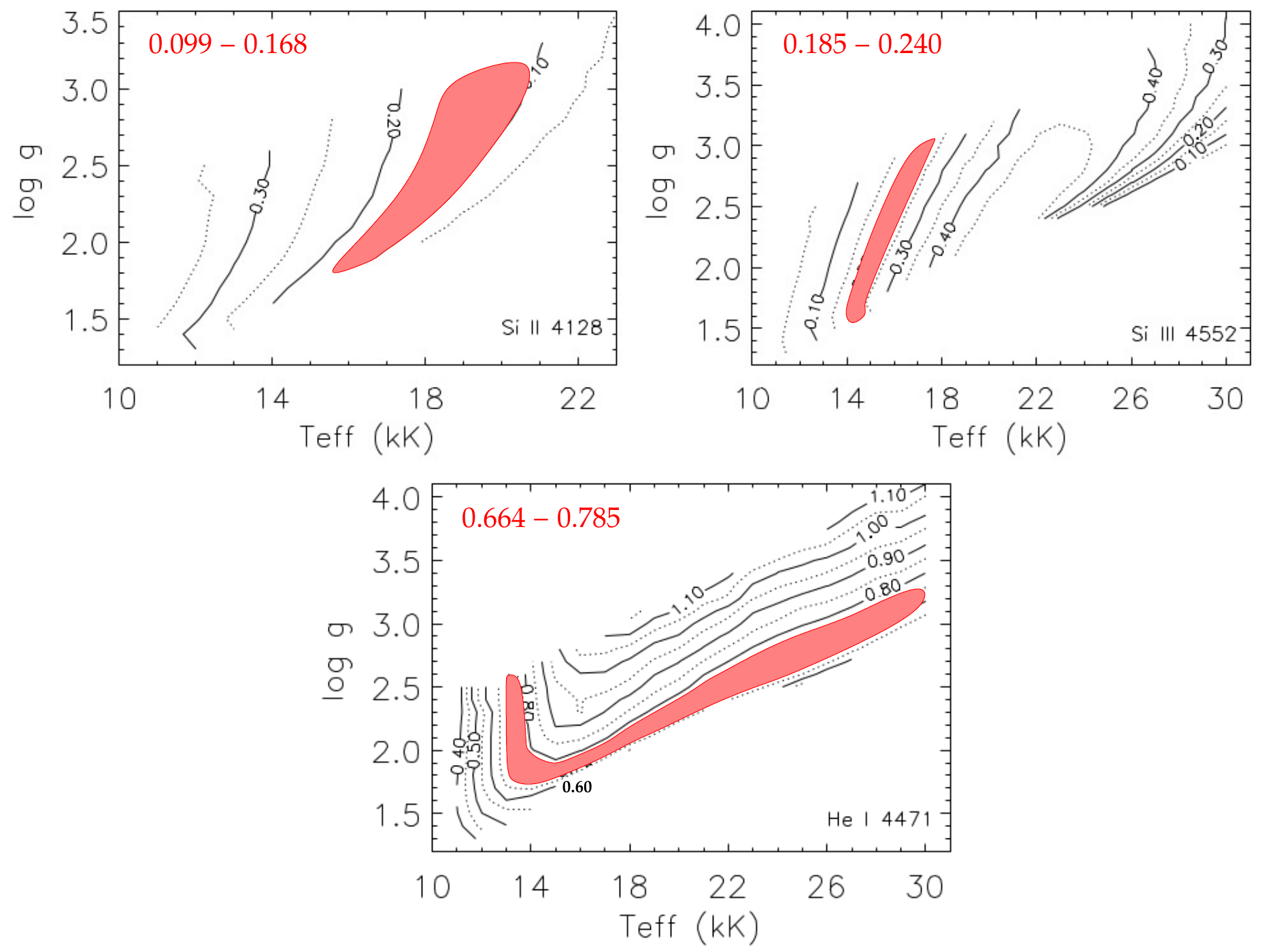} 
    \caption{EW isocontour levels as function of $T_{\rm eff}$ and $\log g$ of three photospheric lines computed by \citet{2007A&A...463.1093L}. The range in EW values covered by our measurements is indicated, and the likely parameter combination is marked by red shaded regions. } 
         \label{fig:Lefever}
\end{figure}

\begin{figure*}
    \centering
    \includegraphics[width=0.8\columnwidth]{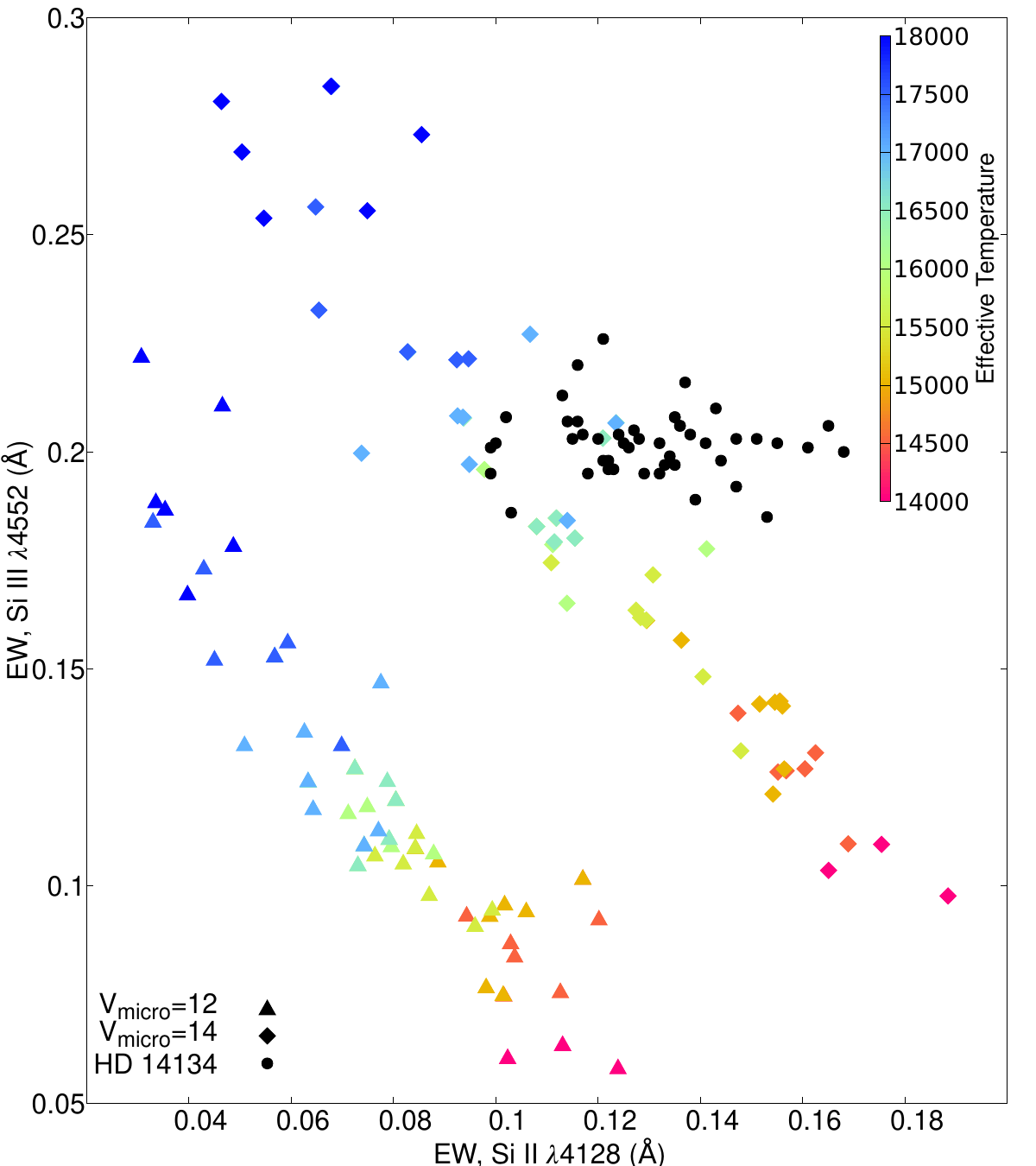} \hspace{1cm}
  \includegraphics[width=0.8\columnwidth]{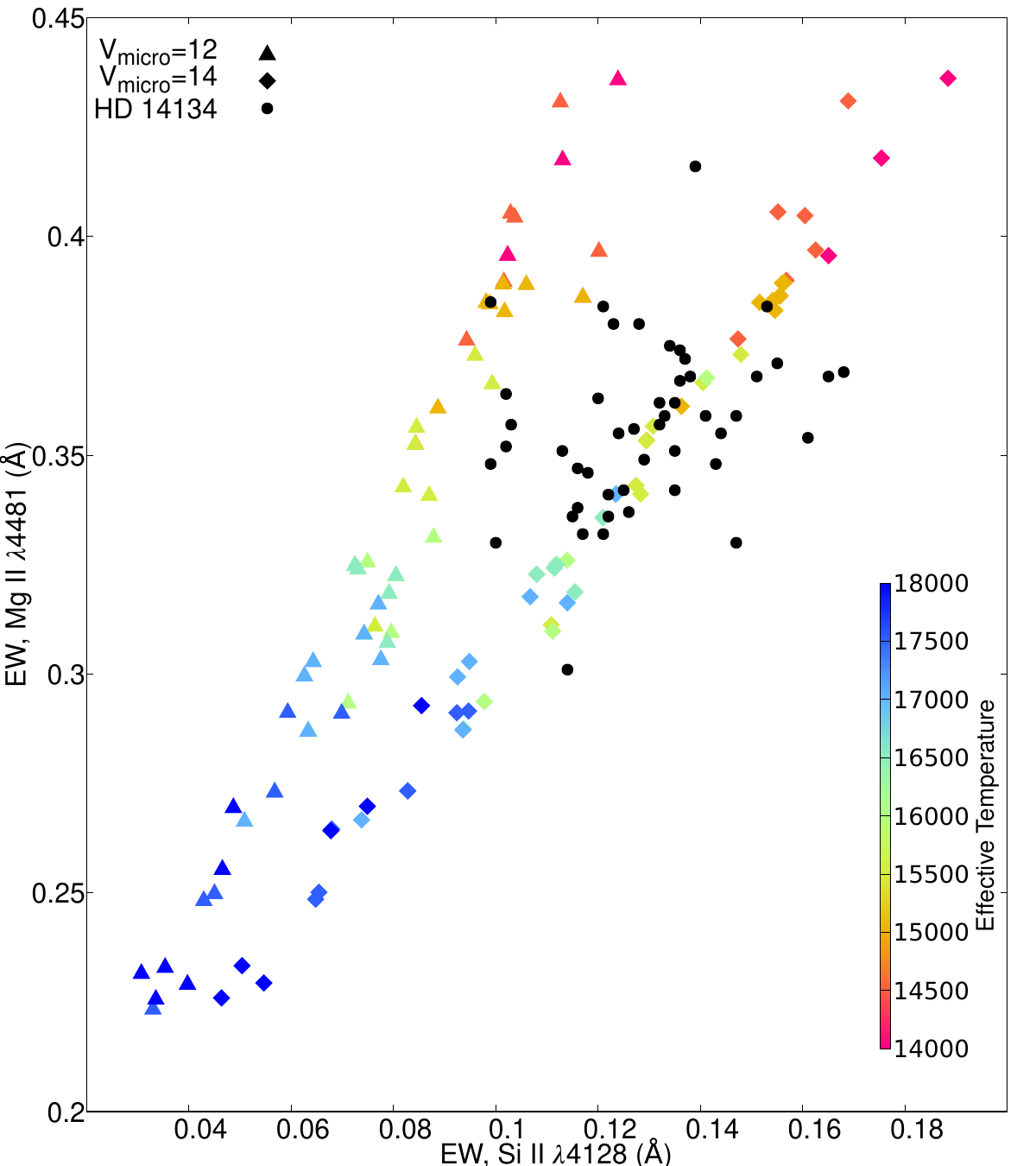}
    \caption{Synthetic EWs for Si\,{\sc iii}\,$\lambda$4552 (left) and Mg\,{\sc ii}\,$\lambda$4481 (right) vs. Si\,{\sc ii}\,$\lambda$4128 for the two values of $v_{\rm mic}$ (different symbols) and for the range of investigated temperatures (colour-coded).
      The measurements for HD\,14134 are shown with black dots.}
    \label{fig:Eqw_model}
\end{figure*}

In Fig.~\ref{fig:EW} we show the values of \citet{1993A&AS...97..559L} for B-type supergiants of luminosity 
class Ia (black circles). These authors also observed HD~14134, and their measurements are highlighted by 
red filled circles. Our measurements are added in blue. The mean values and their standard deviations are 
indicated by solid and dotted lines, respectively. We also include the entire range of measured values, from 
minimum to maximum, which is marked by the dashed lines. Obviously, our individual measurements display a spread 
in values that is considerably higher than typical measurement errors. 

HD~14134 was reported in several studies to be of spectral type B3Ia \citep{1955ApJ...122..429J, 
1992A&AS...94..569L, 2024arXiv240704163N}. However, inspection of the panels in Fig.~\ref{fig:EW} demonstrates 
that the various temperature sensitive lines propose different (or a range of) spectral types\footnote{The 
He\,\textsc{i}\,$\lambda\lambda 4471, 6678$ lines, although included in Table~\ref{tab:Eq_width},
are not used due to little or no variation of their EWs over several spectral sub-types \citep[see Figs.~4 and 
9 in][]{1993A&AS...97..559L}.}. Considering only the mean values and their standard deviation, the lines of 
Si\,{\sc ii} and Si\,{\sc iii} indicate a likely spectral type of B2.5-B3 whereas according to 
Mg\,{\sc ii} it might spread to B4. Most striking is that N\,{\sc ii} suggests a spectral type B5. Inspecting 
the entire range of our measured values, the spread in spectral type (and hence in effective temperature) is 
even larger (from B2 to B6). Also, the measurements by 
\citet{1993A&AS...97..559L} for the lines Si\,{\sc ii}\,$\lambda 6347$ and N\,{\sc ii}\,$\lambda 4601$ fall 
outside our average range, and for the latter line even outside the entire range of our measurements.

As pointed out by \citet{1993A&AS...97..559L}, the comparison with EWs measured from observations (that were 
also collected with lower resolution than our data) might bear some inaccuracies. Therefore, we also 
compared our measurements with predictions from the extensive grid of synthetic models computed by 
\citet{2007A&A...463.1093L} for the temperature range of blue supergiants ($T_{\rm eff} = 10\,000 - 30\,000$\,K) 
and for a range of surface gravities ($\log g = 1.3 - 4.0$ in cgs units). In Fig.~\ref{fig:Lefever} we show the 
isocontour levels of the equivalent widths of the three modeled lines, Si\,{\sc ii}\,$\lambda$4128, 
Si\,{\sc iii}\,$\lambda$4552, and He\,{\sc i}\,$\lambda$4471. These diagrams were taken from the work of 
\citeauthor{2007A&A...463.1093L} (\citeyear{2007A&A...463.1093L}, their Figs.~2 - 4). In each panel, we 
indicate the range of our EWs and mark with red shaded areas the parameter range for $T_{\rm eff}$ and $\log g$ 
suggested by our measurements for these lines. The most likely $T_{\rm eff}$ and $\log g$ combination for our 
star would be at the common intersection of the three ranges. However, we note that He\,{\sc i} has either a 
common intersection region with Si\,{\sc iii} for low temperatures ($T_{\rm eff} = 14\,000-15\,000$\,K) and 
surface gravities ($\log g \sim 1.8-1.9$) or with Si\,{\sc ii} for higher values of temperature ($T_{\rm eff} = 
16\,000-19\,000$\,K) and surface gravity ($\log g \sim 1.8-2.45$). Surprisingly, the ranges of measured EW 
values for the line of Si\,{\sc iii} and Si\,{\sc ii} have no common area. While offsets between the parameters 
suggested from helium and silicon might be explained by different abundances of one (or both) of the elements, 
for which \citet{2007A&A...463.1093L} used solar values, the clear mismatch between the suggested parameter 
combinations from the Si lines might be due to the assumed solar Si abundance. However, we also note that the 
grid of models has been computed for a fixed value of $10$\,km\,s$^{-1}$ for $v_{\rm mic}$, while 
\citet{2024A&A...687A.228D} found a value of $14$\,km\,s$^{-1}$ from their analysis of HD~14134 in agreement
with the general trend of high $v_{\rm mic}$ values in blue supergiants \citep{2018A&A...614A..91H}. A 
higher $v_{\rm mic}$ results in higher values of the EW of the lines and could indeed resolve the 
discrepancy, because the synthetic EWs for Si\,{\sc ii}\,$\lambda$4128 would shift to higher values of 
$T_{\rm eff}$ while those for Si\,{\sc iii}\,$\lambda$4552 would simultaneously shift to lower values. 

To further explore the impact of $v_{\rm mic}$ on the EWs of our strategic lines, we developed a grid of CMFGEN 
models with $v_{\rm mic}$ values of 12 km s$^{-1}$ and 14 km s$^{-1}$. The models span the temperature 
range $T_{\rm eff} = 14\,000 - 18\,000$\,K, have surface gravities around $\log g = 2.0$ (with a spread from 
$1.85 - 2.3$) in cgs units, and are computed for solar abundances. In each synthetic spectrum, we measure the 
EWs of the lines Si\,{\sc ii}\,$\lambda$4128, Si\,{\sc iii}\,$\lambda$4552, and Mg\,{\sc ii}\,$\lambda$4481 
and compare them with our measurements for HD\,14134. In Fig.~\ref{fig:Eqw_model}, we show the EWs for 
Si\,{\sc iii}\,$\lambda$4552 (left) and Mg\,{\sc ii}\,$\lambda$4481 (right) vs. Si\,{\sc ii}\,$\lambda$4128 
for the two values of $v_{\rm mic}$ (different symbols) and for the range of temperatures, which is 
colour-coded. Our measurements for HD\,14134 are depicted with black dots. 

We note that EWs of the Si lines are very sensitive to the value of $v_{\rm mic}$. The synthetic values for the 
investigated temperature range align along straight lines that are parallel in the diagram (left panel of 
Fig.~\ref{fig:Eqw_model}). Comparison with the measurements for HD\,14134 suggests that $v_{\rm mic}$ could on 
average even be slightly larger than 14\,km\,s$^{-1}$. Extrapolation of the synthetic EW values would then result 
in a range of stellar effective temperatures of $15\,000-17\,000$\,K. A similar temperature range is predicted 
from the EW values of the Mg\,{\sc ii} line (right panel of Fig.~\ref{fig:Eqw_model}). Here, the curves for 
constant $v_{\rm mic}$ are not parallel but converge for higher temperatures. A slightly higher value for 
$v_{\rm mic}$ is needed here as well to match with some of the observed values, although most values cluster 
around 14\,km\,s$^{-1}$.

\begin{figure*}
    \centering
\includegraphics[width=0.31\textwidth]{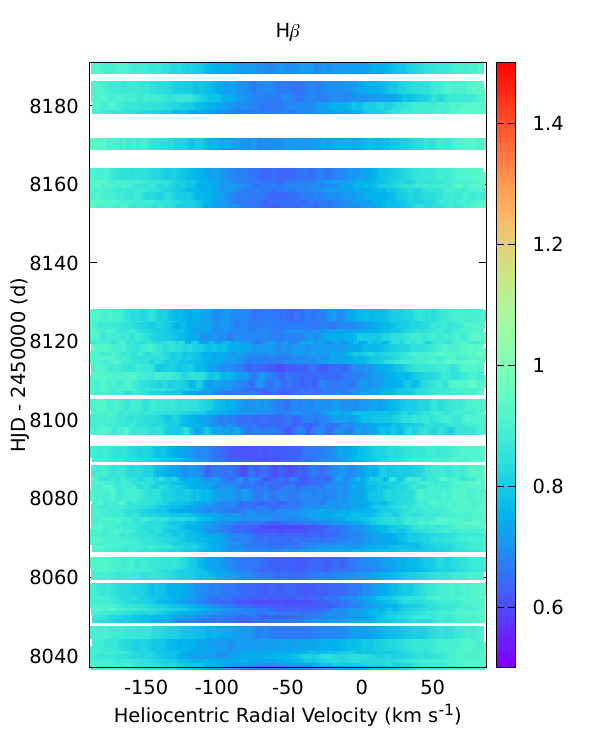}
\includegraphics[width=0.31\textwidth]{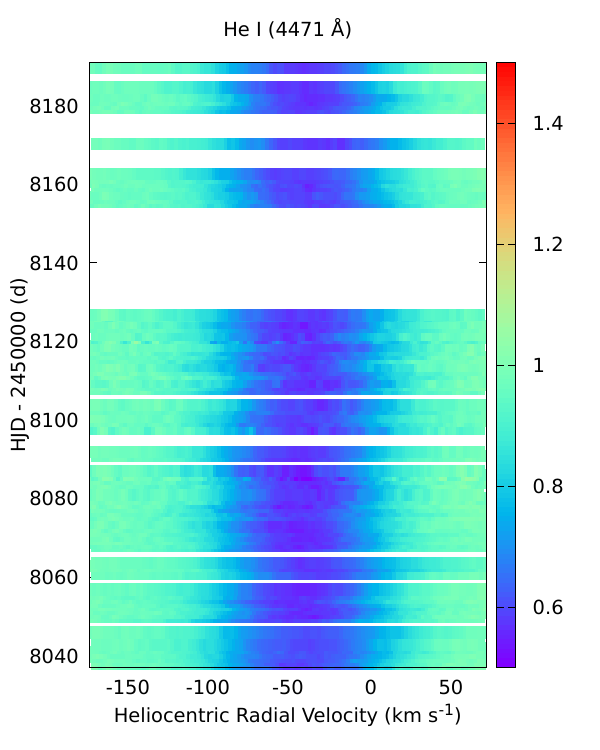}
\includegraphics[width=0.31\textwidth]{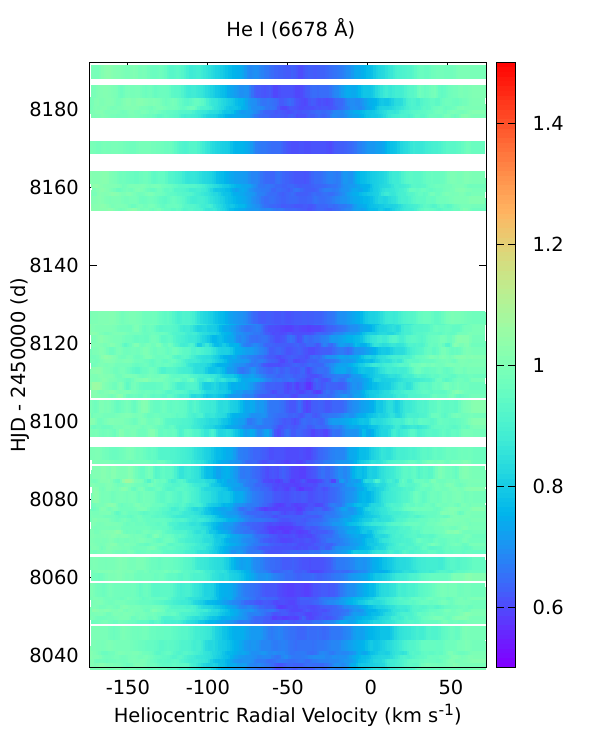}

\includegraphics[width=0.31\textwidth]{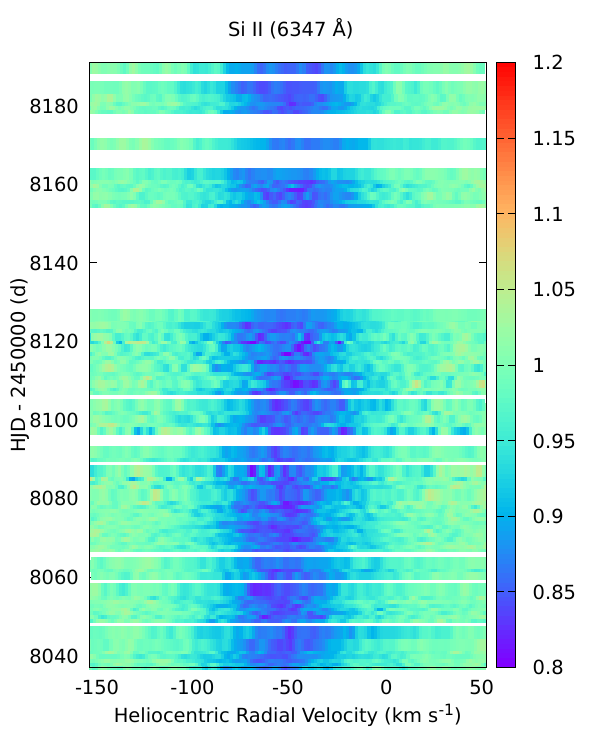}
\includegraphics[width=0.31\textwidth]{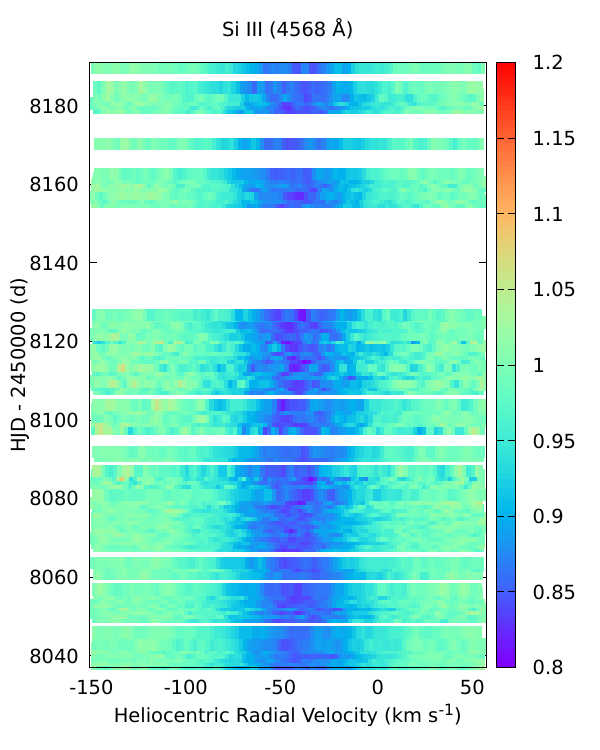}
\includegraphics[width=0.31\textwidth]{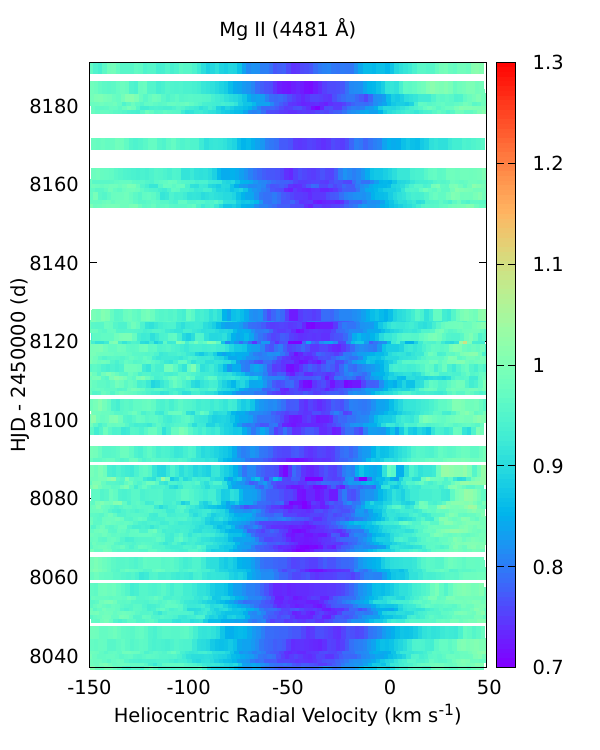}
    \caption{Dynamic plots of selected absorption lines showing the temporal radial velocity variations.}%
    \label{fig:lines}
\end{figure*}

The noticeable spread in the observed EW values, which is significantly larger than the errors of the individual 
measurements, in combination with the rather irregular photometric variability of the star might suggest real 
variations (fluctuations) in stellar parameters\footnote{Although some influence due to imperfect spectral 
normalization cannot be excluded.}, possibly related to a highly dynamical stellar surface (respectively 
stellar atmosphere) due to pulsations\footnote{That pulsations might lead to slight changes in stellar parameters was also recently proposed by \citet{2025A&A...703A...2K} for 
$\zeta$~Oph and shown based on theoretical models for the blue supergiants MWC\,137 \citep{2024MNRAS.527.7414P} and 
$\epsilon$~Ori \citep{2026MNRAS.549ag895D}.}. Atmospheric motions also impact the line-forming regions (in 
particular the depth where the lines are formed), which could explain the observed spread in $v_{\rm mic}$. 
Moreover, they would explain why the EW values of Si\,{\sc iii} are rather stable, whereas those of Si\,{\sc ii} 
show a large scatter. If the variability in EW values would be caused by (pure) temperature variations, then the 
EW values of both lines should vary simultaneously.

\begin{figure*}
    \centering
    \includegraphics[width=0.85\textwidth]{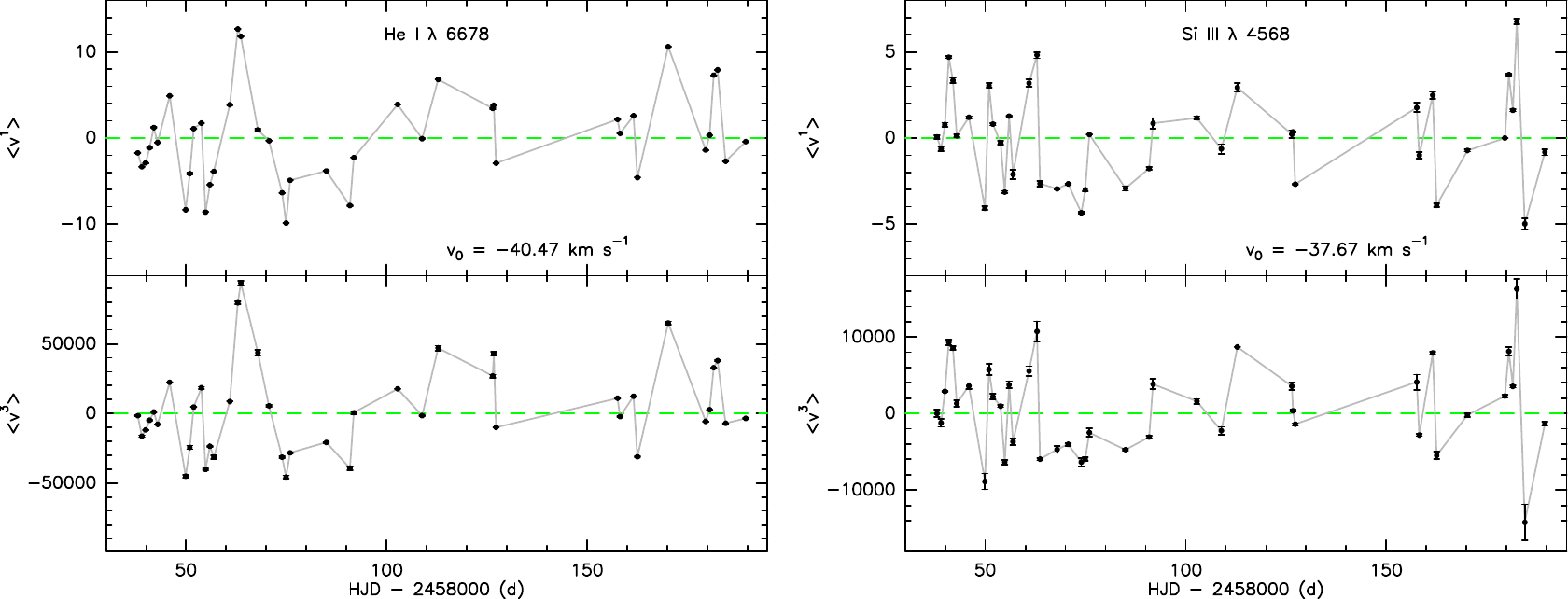}
    \caption{First (radial velocity, top panels) and third moments (line asymmetry, bottom panels) of the He\,{\sc i} $\lambda$6678 (left) and the Si\,{\sc iii} 
    $\lambda4568$ (right) lines.}
    \label{fig:moments}
\end{figure*}

\subsection{Radial velocity variations}

The photospheric absorption lines of HD~14134 display temporal variations in radial velocity. 
Figure~\ref{fig:lines} depicts dynamic plots for the lines H$\beta$, 
He\,\textsc{i}\,$\lambda\lambda$\,4471, 6678, 
Si\,\textsc{ii}\,$\lambda$\,6347, 
Si\,\textsc{iii}\,$\lambda$\,4568 and 
Mg\,\textsc{ii}\,$\lambda$\,4481.
The colour code represents the flux normalized to the continuum value. The profiles change clearly in width, 
depth, and radial velocity. In many cases, we also observe asymmetric profile shapes. But no obvious periodic 
variability is seen. The asymmetry noticeable in the H$\beta$ line, in particular at the beginning of the time 
series, is caused by wind emission superimposed on the photospheric absorption line. Although 
the diverse spectral lines are formed in different layers of the stellar atmosphere, the patterns of radial 
velocity variability are quite similar.

The object HD~14134 was included in the recent study by \citet{2024arXiv240511209S}, who analysed the line 
Si\,{\sc iii} $\lambda$ 4568 and identified line-profile variability with a peak-to-peak amplitude in radial 
velocity $\le 5$\,km\,s$^{-1}$. To test whether this line-profile variation could be due to pulsation activity, 
we selected the same line and computed its moments \citep{1992A&A...266..294A, 1994A&A...288..155N}. We limited our 
analysis to spectra with S/N values $\ge 70$ and integrated the fluxes below 0.98 in the normalized spectra. We 
are aware that for precise results from the moment method higher quality data (S/N $\sim 300$) are required, but
our intention is not to determine pulsation modes, but simply to check the global behaviour of the moments. For 
this, we compared in the right panel of Fig.~\ref{fig:moments} the temporal variation of the first (radial 
velocity, top panel) and third (line asymmetry, bottom panel) moments. These were corrected for the 
presumable radial velocity of the star (indicated in the top panel). From these plots, we note that the general 
behaviour of the two moments is very similar. We did the same analysis for the He\,{\sc i} $\lambda 6678$ line, 
and show the results on the left side of Fig.~\ref{fig:moments}. Also for this line, the two moments vary in 
phase and show a similar trend over the entire observing period as the Si\,{\sc iii} $\lambda$ 4568 line. This 
is a strong indication for pulsations as cause for the variability rather than other effects, such as stellar 
spots. But we also see some differences between the two elements, which we interpret as due to wind variability 
that can have some influence on the profile of the He\,{\sc i} $\lambda 6678$ line. The amplitude of the radial 
velocity variation of the two lines is quite different. While a peak-to-peak amplitude of $\sim 5$\,km\,s$^{-1}$ 
is found from Si\,{\sc iii} (with an exception of $\sim 7$\,km\,s$^{-1}$ towards the end of our observing run), 
which is similar to what was found by \citet{2024arXiv240511209S}, the velocity amplitude of He\,{\sc i} is more 
than twice as large.

\subsection{Stellar wind variability}
\label{sect:wind}

The winds of mid- to late-type B supergiants imprint their signals mainly on the hydrogen Balmer lines, 
especially on H$\alpha$. HD\,14134 is known for its highly variable H$\alpha$ profile with pure absorption,  P~Cyg profiles, as well as complete absence of H$\alpha$ likely due to 
compensation of the stellar photospheric absorption line with wind emission \citep{2004MNRAS.351..552M, 
2017RAA....17...38M, 2017JApA...38...20M}.

\begin{figure}[ht!]
    \centering
        \includegraphics[height=7cm]{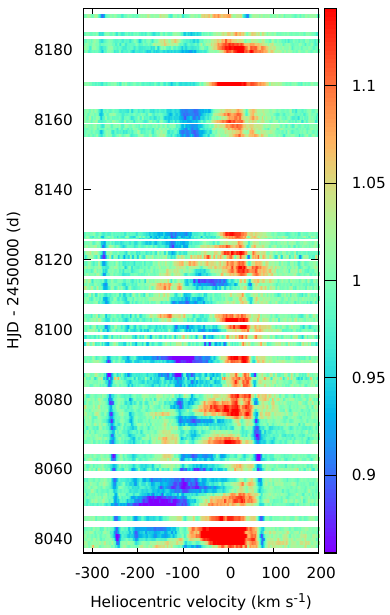}~
        \includegraphics[height=7cm]{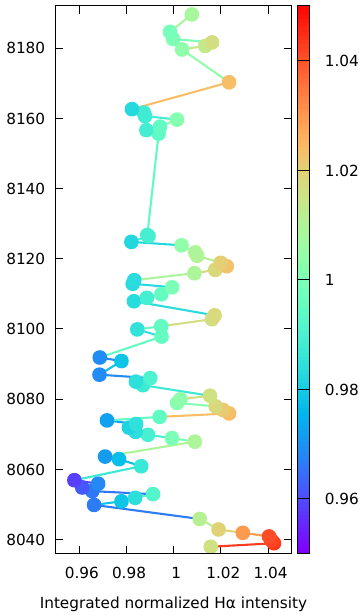}
    \caption{Left: Dynamic plot showing the temporal evolution of the H$\alpha$ profile. The sharp absorption 
    features with radial velocity drift are telluric lines. Right: normalized intensity integrated over the 
    H$\alpha$ line. The colour bar gives the normalized intensity. 
}
    \label{fig:halfa}
\end{figure}

Our spectra collected over a period of five months reflect this diversity (see Fig.~\ref{fig:Ha_profile}). In 
the left panel of Fig.~\ref{fig:halfa} we present a dynamic plot of the H$\alpha$ line. 
Obviously, H$\alpha$ had an intense and broad emission and a shallow absorption component at the beginning of 
our monitoring. Over the course of the observations, the absorption component shows strong variability in radial 
velocity, width, and intensity. At several dates the absorption and/or emission components are very weak or even 
disappear (almost completely) from the spectrum. At some phases, the emission displays an additional blue peak 
mimicking a double-peaked emission profile as is known for stars with disks \citep[such as Be stars, see 
e.g.][]{2003PASP..115.1153P}, and the profile eventually develops further into an inverse P Cygni, which might 
indicate that some material is falling back to the stellar surface.

No clear periodicity is found in the evolution of the 
H$\alpha$ profile, but the strong variability in the 
emission and absorption components of H$\alpha$ raises the question whether it can (significantly) influence 
the observed photometry. While both Gaia photometers cover the wavelength region of H$\alpha$  
\citep{2018A&A...616A...4E}, the colour (bottom panel of Fig.~\ref{fig:gaia}) should be uninfluenced by the 
H$\alpha$ line variability, and the colour changes of about 0.1\,mag might trace temperature variability of the 
star\footnote{Unfortunately, no effective temperature calibration for BSGs based on Gaia colours exist to date, 
so that we cannot quantify the effective temperature of HD\,14134 and its temporal changes based on the Gaia 
photometric measurements.}. The TESS detector bandpass is centred on the traditional Cousins $I$-band and spans 
from $6000$\,\AA \ to $1\,\mu$m including H$\alpha$. Therefore, to visualize the effect the time-variable wind 
of HD~14134 might imprint on photometric measurements, we integrated the spectra over the wavelength range 
$6557-6567.5$\,\AA \ and normalized them
\begin{equation}   
F_{\rm norm} = \frac{\int_{\lambda_{1}}^{\lambda_{2}} F_{\lambda} \, d\lambda}{\int_{\lambda_{1}}^{\lambda_{2}}
d\lambda }\, .
\end{equation}
The results are shown in the right panel of Fig.~\ref{fig:halfa}. During the enhanced phase of wind activity at the 
beginning of the monitoring, the variability exceeds $4\%$, whereas during more quiescent phases it varies 
around $2-3\%$. The amplitude of the photometric variability seen in the TESS light curves (Fig.\,\ref{fig:lc}) 
is of the same order of magnitude. This suggests that most of the photometric 
variability observed by TESS might be caused by the time-variable wind and that the variability due to pulsations
might play only a minor role. We return to this in Sect.\,\ref{sect:puls_wind}.

\section{Analyses of the light curves}\label{sect:photometry}

Although the wind seems to have major impact on the light curve, the variability seen in the photospheric
lines suggests the presence of pulsation activity. The sparse sampling of the Gaia data 
(Fig.~\ref{fig:gaia}) is unsuitable to reliably identify periodic signals. Therefore, we focus on the TESS 
light curves (Fig.~\ref{fig:lc}), which display prominent but irregular variability along with some inherent 
variations on time-scales considerably longer than the sector length. In the following, we analyse the light 
curves to search for signatures that might be related to pulsations.

\subsection{Fourier analysis}
\label{sect:Fourier}

\begin{figure}
    \centering
    \includegraphics[width=\columnwidth]{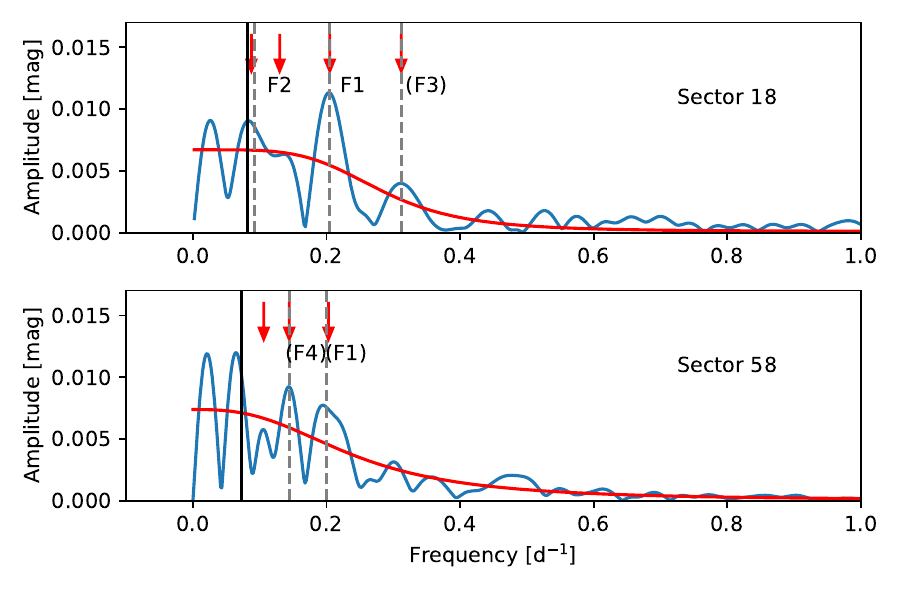}
    \caption{Amplitude spectra of the first Fourier transform of the TESS light curves from sector 18 (top) 
    and 58 (bottom). Red curves represent the best-fit models for the Lorentzian function given by Eq.\,(\ref{Eq.rednoise}). Vertical black lines 
    depict the lower limit for confident frequency detection. Grey dashed lines and red arrows
    mark frequencies identified using \textsc{Period04} and the Wavelet analysis (see Sect.~\ref{sect:wavelet}), respectively.}
    \label{fig:enter-Fourier}
\end{figure}

We started in the classical way to retrieve the frequencies of possible periodic signals and used
\textsc{Period04} \citep{2005CoAst.146...53L} to compute the discrete Fourier Transforms for non-uniformly sampled data.
Figure~\ref{fig:enter-Fourier} shows the First Fourier Transform of the TESS data from both sectors.
Our analysis is confined to the interval [0-24]\,d$^{-1}$, below the Nyquist frequency of each sector. We found 
no strong signal beyond 1\,d$^{-1}$. The amplitude and phase are calculated using a least square sine fit for 
each detected frequency. After obtaining the first frequency, the analysis is performed on the residuals in a 
standard pre-whitening procedure. The uncoupled uncertainties in the frequencies and amplitudes are calculated 
using Monte Carlo simulations, following the procedure outlined in \citet{2023A&A...676A..96S}. 

To select the most likely frequencies from the obtained list, we applied the following criteria. We adopted a 
conservative S/N $\ge$ 5, for each sector, to avoid spurious peak detection induced by noise, following the 
recommended values in \cite{2021AcA....71..113B}. For the computation of the S/N value of each frequency, we 
considered a spectral window size of 1~d$^{-1}$. As was shown by \citet{2020A&A...639A..81B}, such a narrow window 
size prevents from over-interpretation of the data for stars which have prominent 
SLF variability as seems to be the case for our star. We further adopted as resolution criteria 
$1.5/\Delta T$ as in \citep{1978Ap&SS..56..285L}, being  $\Delta T$  the observation time span of the 
considered sector. For sectors 18 and 58 the time span is 24.33\,d and 27.71\,d, resulting in values of 
0.0616\,d$^{-1}$ and 0.0541\,d$^{-1}$, respectively, for the minimum separation distance of neighbouring 
frequencies. When comparing frequencies from different sectors, we adopted a conservative frequency separation of 
0.05411\,d$^{-1}$. A further constraint is that for a confident identification, the time span of the
observations should spread over at least twice the period. This means that we considered only frequencies greater 
than $2/\Delta T$, i.e., greater than 0.0822\,d$^{-1}$ and 0.0722\,d$^{-1}$ for sectors 18 and 58, respectively.

The final frequencies are listed in Table~\ref{tab:FFT}. Only two significant frequencies are obtained in 
sector 18, whereas the first two highest amplitude frequencies derived in sector 58 fall below our S/N 
threshold and are, therefore, not considered as reliable identifications. But we list them, because 
(F1) seems to agree with F1 found in sector 18, although it displays a significantly lower S/N. And (F4) is 
listed because it has a higher amplitude than (F1). Similarly, we list (F3) in sector 18. All these frequencies 
seem to be affected by the high level of the underlying noise.

To quantify this (red plus white) noise, we followed the approach of \citet{2019A&A...621A.135B} and fitted the 
amplitude spectra in Fig.\,\ref{fig:enter-Fourier} with a Lorentzian function of the form
\begin{equation}
\label{Eq.rednoise}
    \alpha(\nu)= \frac{\alpha_0}{1+(\frac{\nu}{\nu_{\rm char}})^\gamma} + C_{\rm W},
\end{equation}
where $\alpha_0$ defines the amplitude at frequency zero, $\gamma$ is the gradient of the linear part of the 
profile in a logarithmic presentation, $\nu_{\rm char}$ is a characteristic frequency representing the inverse 
of the mean duration of the dominant structures in the light curve, and $C_{\rm W}$ is the frequency-independent 
level of the white noise. To derive the best-fitting parameters, we performed a Monte Carlo simulation in the 
frequency range [0-24] d$^{-1}$. Specifically, we calculated the standard deviations of the residuals after 
fitting our data with the Python package \texttt{curve fit} \citep{Vugrin} and used them to create synthetic Gaussian noise. The 
noise was randomly generated in each iteration and added to the original data, creating the synthetic sample for 
each iteration. Finally, we computed the mean and the standard deviation after fitting Eq.\,(\ref{Eq.rednoise}) 
to our synthetic data 1000 times. The results are displayed in Table \ref{table:red-noise} and the fit curves
are included in Fig. \ref{fig:enter-Fourier}. We notice that the parameters considered to model the noise 
are different for the two sectors. Changes in the behaviour of SLF variability might indicate that its origin is 
linked to short-term variations that do not prevail from one sector to another and might be related (at least 
partially) to the previously identified wind variability. The drop in the level of white noise, $C_{W}$, by 
almost a factor of three is most likely due to the different cadences in the studied sectors.

\begin{table}
\centering    
\caption{Best fit parameters for the Lorentzian functions (Eq.\,\ref{Eq.rednoise}).}         
\label{table:red-noise}
\begin{tabular}{ccc}        
\hline\hline  
Parameters & Sector 18 & Sector 58 \\
\hline                      
  $\alpha_0$ [$\mu$mag]& 6613.23 $\pm$ 44.28 & 735.31 $\pm$ 49.51 \\ 
   $\nu_{\rm char}$ [d$^{-1}$] & 0.282 $\pm$ 0.001 & 0.239 $\pm$ 0.002 \\
   $\gamma$ & 4.52 $\pm$ 0.29 &  2.74 $\pm$ 0.06 \\ 
   $C_{\rm W}$  [$\mu$mag] & 99.13 $\pm$ 3.36 & 37.63 $\pm$ 2.78 \\ 
   \hline
 \end{tabular}
\end{table}

\subsection{Wavelet analysis}
\label{sect:wavelet}

\begin{figure}
    \centering
    \includegraphics[width=0.9\columnwidth]{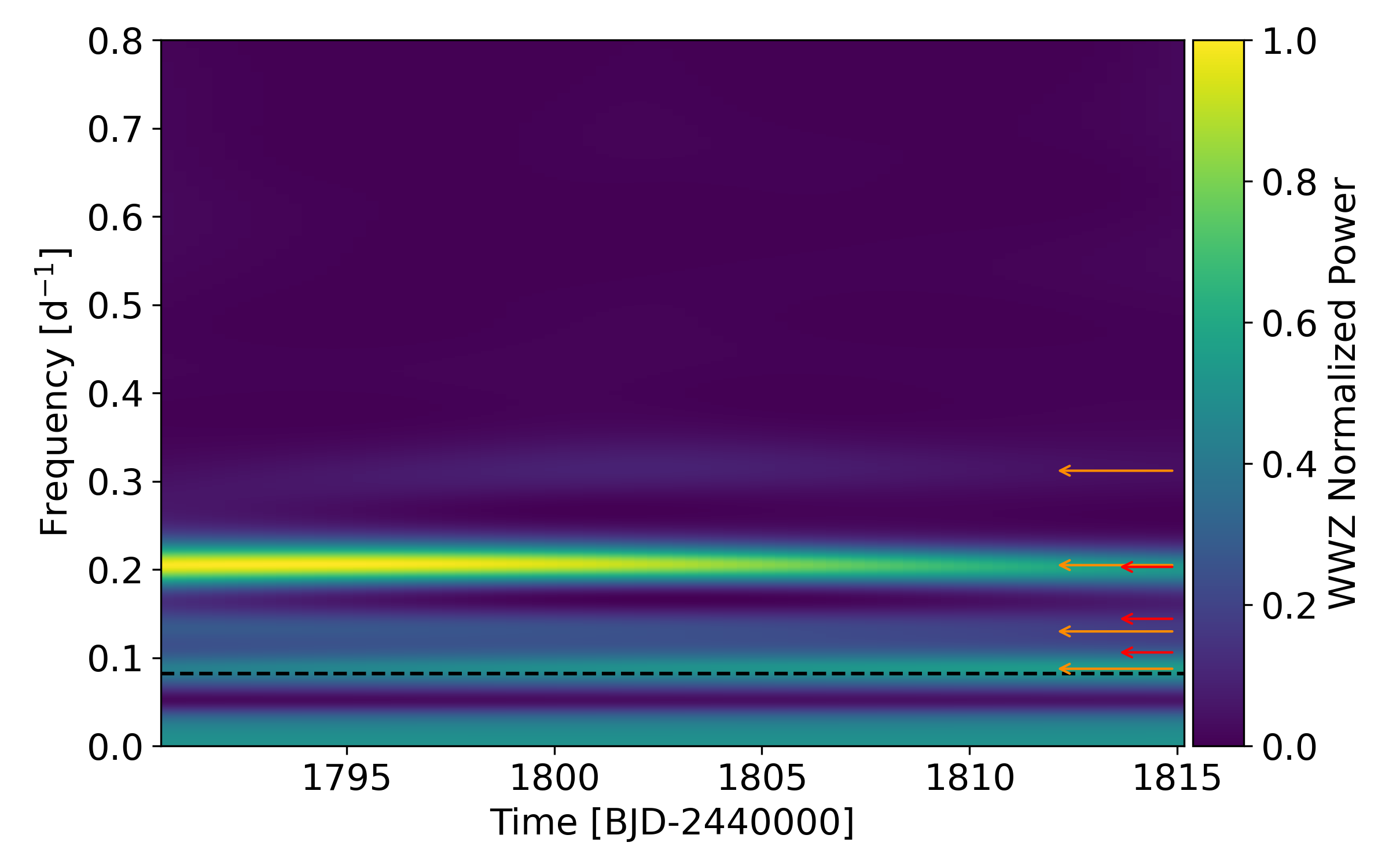}
    \includegraphics[width=0.9\columnwidth]{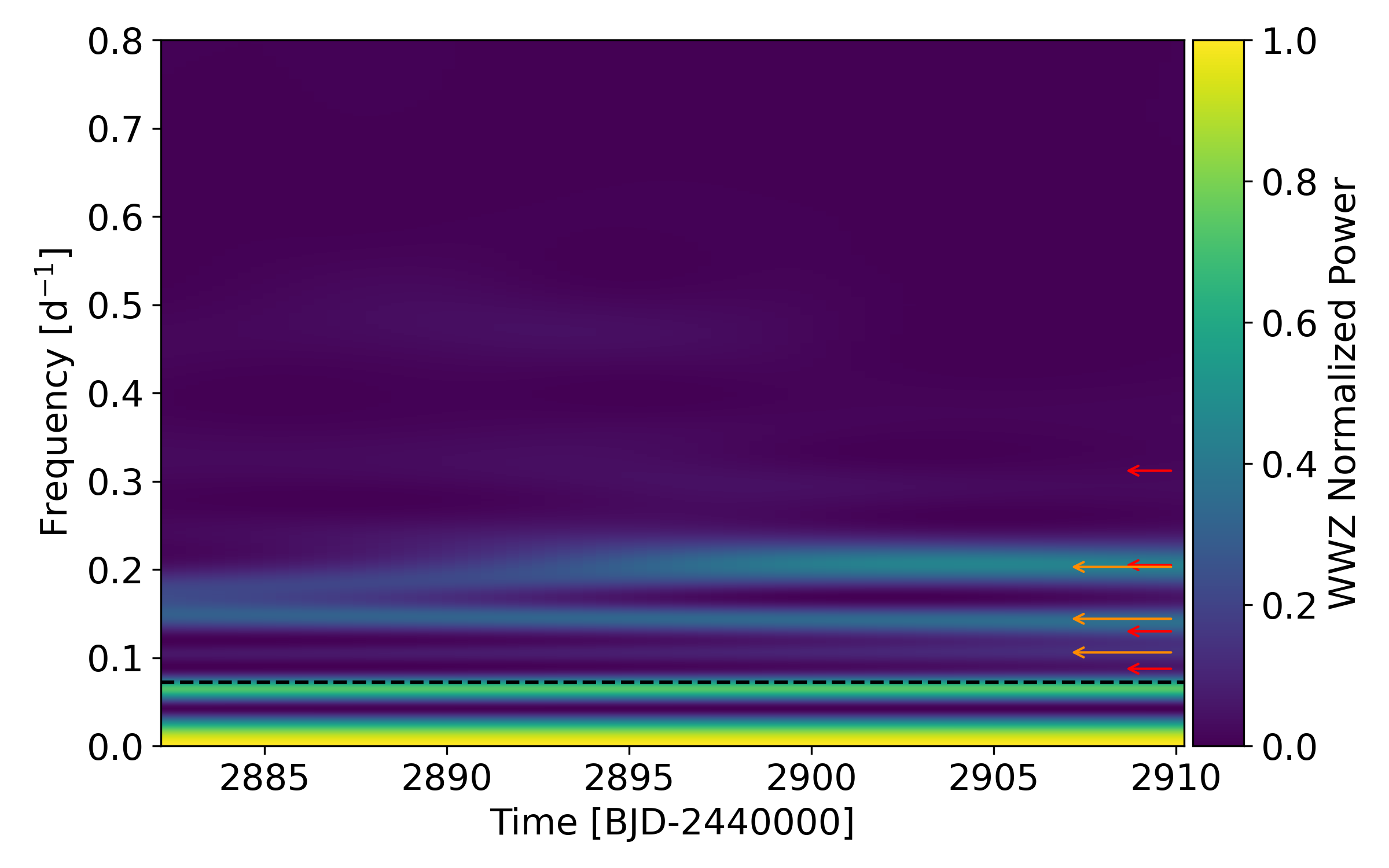}
    \caption{Scalograms for the light curves of sector 18 (top) and 58 (bottom). Orange (red) arrows indicate 
    the frequencies obtained from the averaged power resulting from the scalogram of the same (other) sector. 
    The black dashed lines mark the lower limit for reliable frequency identification in each sector.}
    \label{fig:WWZ}
\end{figure}

Our Fourier analysis shows that the frequency with the highest amplitude, F1, identified in sector 18 is barely 
detectable in sector 58. F2 does not prevail over time and is not detected in sector 58. Similarly, (F4) which 
has the highest amplitude in sector 58 has no counterpart in sector 18. This suggests that the defining 
parameters of the signals (frequency, amplitude, phase) in the light curves are time variable. To study the 
changes in the frequencies and their amplitudes over the observing periods, we performed a wavelet analysis, 
more specifically, we employed the Weighted Wavelet Z-Transform \citep[WWZ,][]{1996AJ....112.1709F}. In contrast 
to the classical wavelet method, WWZ is suitable to study unevenly sampled time series. It utilises a revised 
adaptation of the standard Morlet wavelet \citep{grossmann1984decomposition} to generate the local power 
spectra\footnote{The tool we used computes normalized power spectra instead of amplitude spectra, so that we use this notation throughout the paper when referring to WWZ results.}, i.e. the energy-distribution of the signal in the time-frequency domain. 
Figure~\ref{fig:WWZ} represents the WWZ-scalograms obtained from the analysis of the data of both sectors using 
the Python package \texttt{libwwz}\footnote{\url{https://github.com/ISLA-UH/libwwz}}.

The frequencies with the highest power obtained by the time integration of the local power spectra are listed in 
Table~\ref{tab:wwz_freq} using lowercase letters but following the same numbering as before. These frequencies 
agree fairly well with those obtained from \textsc{Period04} (Table~\ref{tab:FFT}). Deviations between the 
two are below 0.052 d$^{-1}$ and are due to the different algorithms employed in each method. Frequencies f4 in 
sector 18 and f5 in sector 58 do not satisfy our selection criteria, however, they are clearly present in the 
amplitude spectrum (see Fig.~\ref{fig:enter-Fourier}). The scalograms demonstrate a decrease in amplitude of the 
main frequency, f1, in sector 18, whereas it is not present at the beginning of sector 58, but appears and 
strengthens during the second half of that observing period. Frequency f2 seems to strengthen over sector 18, 
whereas in sector 58, it has no clear counterpart. It could have either turned into the signal seen at lower
frequency (below the dashed line) but with considerably higher amplitude, or it could have shifted to a slightly 
higher frequency (identified as f5 in sector 58) but with significantly lower amplitude. Moreover, the 
frequencies f1 and f4 in sector 58 seem to be connected before the start of the observing run for sector 58 and 
separate shortly thereafter, with one component gradually converging towards f1, and notably matching the 
dominant frequency observed in sector 18. We also detect a broad but faint feature with time-variable intensity 
in sector 18 labelled as f3, which lacks a counterpart in sector 58.

\begin{figure*}
    \centering
\includegraphics[width=0.32\textwidth]{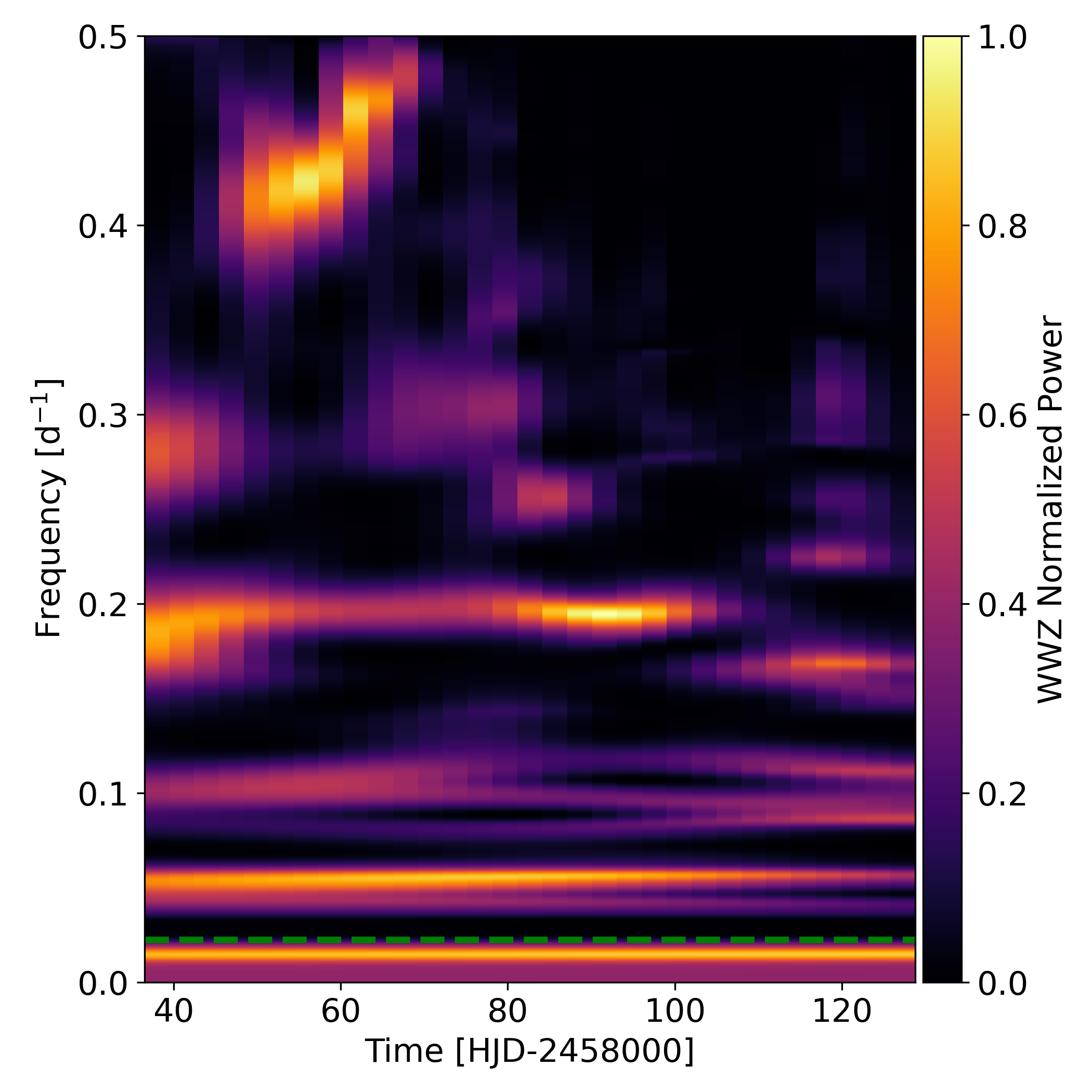}
\includegraphics[width=0.32\textwidth]{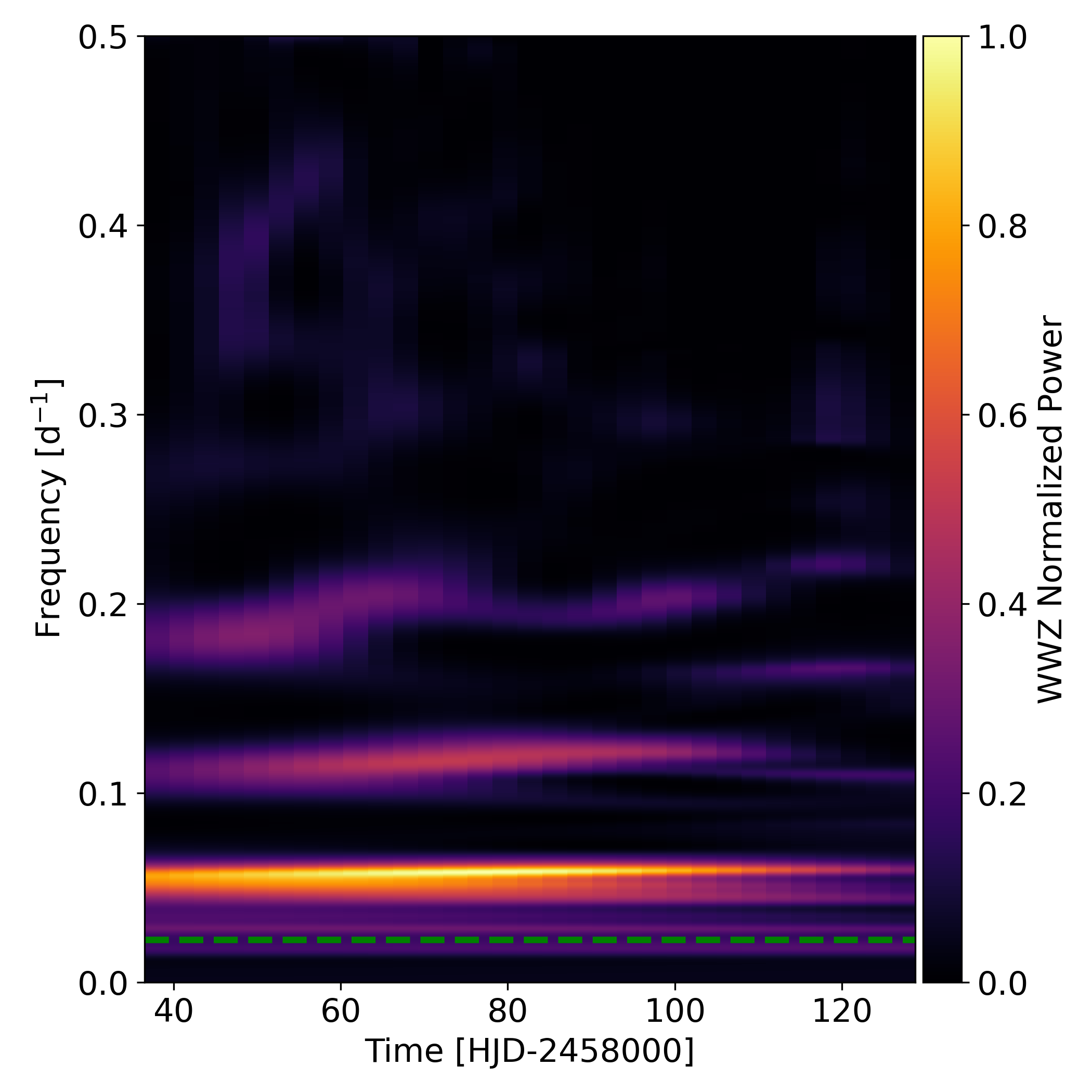}
\includegraphics[width=0.32\textwidth]{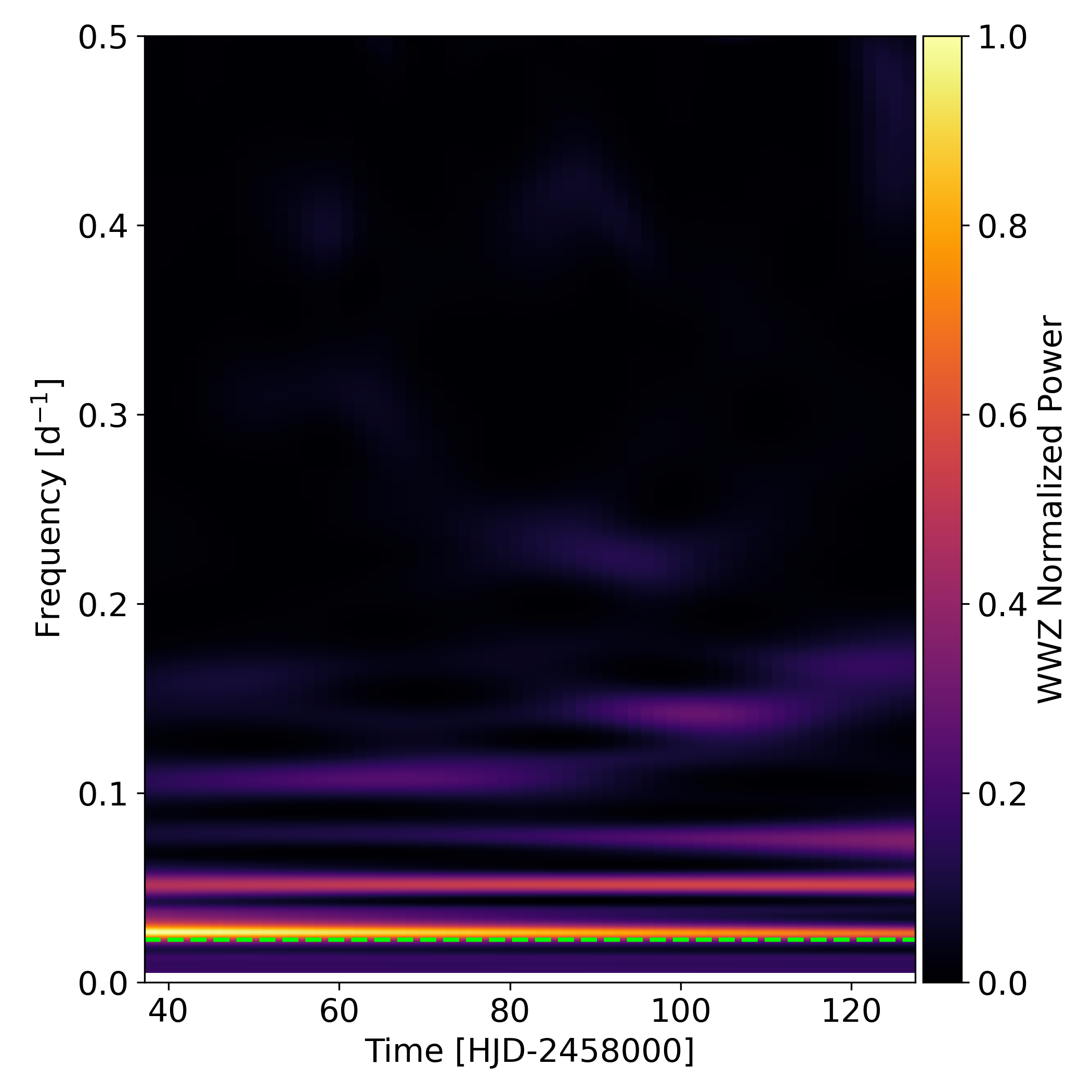}
    \caption{Scalograms of the radial velocity variation of the Si\,{\sc iii} (left) and He\,{\sc i} lines 
    (middle), and of the normalized integrated H$\alpha$ line (right). The horizontal green dashed lines 
    indicate the threshold for the lowest confident frequency identification.}  
    \label{fig:wind_wwz}
\end{figure*}

\begin{table*}[ht!]
    \centering
    {\small
    \caption{Left: List of the frequencies obtained from the scalograms of the radial velocity variation of the 
    Si\,{\sc iii} and He\,{\sc i} lines, the normalized integrated H$\alpha$ line, and the TESS photometry. 
    For the latter, the entries for f1 and f4 in Table~\ref{tab:wwz_freq} were averaged. Middle: Modes predicted by GYRE for the two stellar models in Table\,\ref{tab:mesa_models} with their frequencies and harmonic degree. Right: Possible interpretation.}
    \label{tab:freq_mix}
    \begin{tabular}{cccc|cc|c}
    \hline
    \hline
    \multicolumn{4}{c|}{Frequencies [d$^{-1}$]} & M1 & M2 & Possible\\
      Si\,{\sc iii} & He\,{\sc i} & H$\alpha$  & TESS  & f [d$^{-1}$] (l) & f [d$^{-1}$] (l) & Interpretation \\
      \hline
 (--)                & (--)                & (--)                & (--)                & 0.0185 (1) & 0.0186 (1)  & -- \\
 --                  & 0.0284 $\pm$ 0.0261 & 0.0263 $\pm$ 0.0040 & (--)                & 0.0295 (1)      & 0.0303 (1) & wind ?\\
  --                  & --                  & --                  & (--)                & 0.0452 (3) & -- & -- \\
 0.0559 $\pm$ 0.0058 & 0.0585 $\pm$ 0.0094 & 0.0515 $\pm$ 0.0045 & (--)                & -- & 0.0521 (2) & g-mode, $f_{\rm g}$ \\
 --                  & --                  & 0.0768 $\pm$ 0.0063 & 0.0875 $\pm$ 0.0227 & -- & -- & wind ?\\
 0.0978 $\pm$ 0.0197 & 0.1191 $\pm$ 0.0106 & 0.1076 $\pm$ 0.0115 & 0.1059 $\pm$ 0.0747 & --        & -- & $2\times f_{\rm g}$ \\
         --          &       --            & 0.1428 $\pm$ 0.0207 & 0.1368 $\pm$ 0.0586 & --        & -- & wind \\
 0.1680 $\pm$ 0.0298 & 0.1658 $\pm$ 0.0344 & 0.1663 $\pm$ 0.0436 &                 --  & --        & -- & $3\times f_{\rm g}$ \\
 0.1953 $\pm$ 0.0111 & 0.1981 $\pm$ 0.0275 & 0.2276 $\pm$ 0.0221 & 0.2040 $\pm$ 0.0412 & --        & -- & $4\times f_{\rm g}$ \\
 0.2922 $\pm$ 0.0394 & 0.2931 $\pm$ 0.0806 &       --            & 0.3119 $\pm$ 0.0289 & --        & -- & -- \\
 0.4249 $\pm$ 0.0514 & 0.3278 $\pm$ 0.1447 & 0.4064 $\pm$ 0.0337 &     --              & --        & --  & -- \\
      \hline
    \end{tabular}
    \tablefoot{The entries (--) mark ranges below the lower limit for reliable frequency identifications; -- means no frequency detected in the observations respectively found from the theoretical models.}
    }
\end{table*}

\section{Discussion and Conclusions}\label{sect:discussion}

The study presented here leverages high-cadence space-based TESS photometric light curves and (albeit not 
simultaneously acquired) ground-based spectroscopic time-series to probe the observed variability in HD~14134 
and to constrain the evolutionary state of the object. With its reported spectroscopic and photometric variability, the star was proposed to belong to the $\alpha$~Cyg variables \citep{2017A&A...598A.108L}. 
Such objects display small-amplitude radial velocity and brightness variations 
\citep{1998A&AS..128..117V, 2024SASS...43..145A} that can display alternating phases of 
quasi-periodic and erratic variability as was reported by \citet{2026arXiv260302421G} for the prototype of this 
class, the star $\alpha$ Cygni (Deneb). Studies of the pulsation properties of $\alpha$~Cyg variables suggested 
that these objects might be post-red supergiant stars \citep{2013MNRAS.433.1246S}. In the following, we 
discuss the plausible cause of the detected variabilities in HD~14134 and how our results may help to unveil the nature of the object.

\subsection{Pulsations versus wind variability}
\label{sect:puls_wind}

HD~14134 displays a variety of variabilities: in EW, radial velocities, stellar wind activity, and brightness. On the one hand, the moment analysis of the photospheric absorption lines of Si\,{\sc iii} and He\,{\sc i} suggests that
pulsations might cause the observed radial velocity and line-profile variability. The high value of 
macroturbulence needed to reproduce the line broadening is proposed to indicate pulsation activity, as was also reported for the BSG $\rho$~Leo \citep{2018MNRAS.476.1234A}. The EW 
measurements of temperature-sensitive lines display a large spread that might be interpreted with temperature 
fluctuations within the line-forming regions, likely due to a highly dynamical atmosphere. This again might 
indicate that pulsation activity is the main cause of the observed variations. On the other hand, the analysis of the 
TESS light curves provides no clear evidence of stable, coherent pulsation modes. Instead, the amplitude spectra 
are dominated by SLF variability, hampering a confident identification of periodic signals, and the identified signals 
show temporal variations in either frequency, amplitude, or both. Furthermore, the H$\alpha$ line displays strong 
temporal line-profile variability due to time-variable winds that cause brightness variations on the same order 
of magnitude as the variations detected in the TESS light curves. 

\begin{figure}[ht!]
    \centering
    \includegraphics[width=0.8\linewidth]{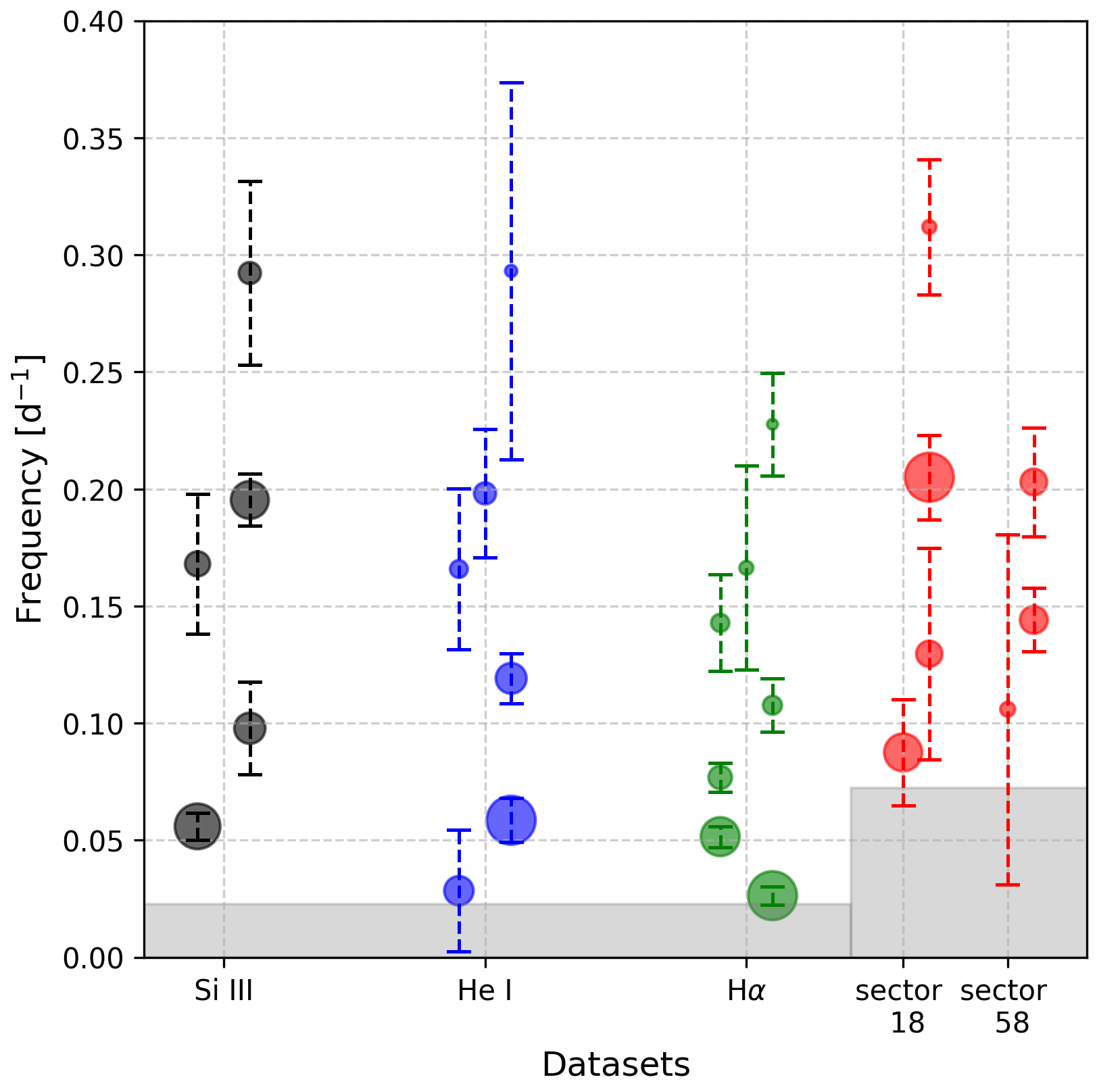}
    \caption{Graphical comparison of the frequencies and their errors (Table~\ref{tab:freq_mix}) obtained from 
    the WWZ analysis of the different observational tracers. The frequencies from both TESS sectors 
    (Table~\ref{tab:wwz_freq}) are shown. The grey box indicates the thresholds. The symbol 
    sizes scale with the power of the signals, which are normalized to the maximum value.}
    \label{fig:freq-mix}
\end{figure}

To explore whether the different observed variabilities might be pulsation-induced or wind-induced, we performed 
a wavelet analysis of the radial velocity curves of the Si\,{\sc iii} and He\,{\sc i} lines as well as of the 
integrated H$\alpha$ line using the WWZ method. We restricted the analysis to the spectra from the first three 
months to avoid artefacts in the scalograms due to the about one month gap in the data. The resulting signals, 
normalized to their respective maximum power, are shown in Fig.~\ref{fig:wind_wwz}. No significant signals are 
found for frequencies greater than $0.5$\,d$^{-1}$. The dashed line in each panel marks the value of
0.0225\,d$^{-1}$. This value indicates the confident identification threshold considering the observation time 
span of $\Delta T= 88.78$\,d.

The scalograms reveal a different frequency pattern compared to the TESS 
data (Fig.~\ref{fig:WWZ}). Three categories appear in certain frequency domains and 
are present in all three data sets: (i) rather stable (or mildly variable in frequency and/or amplitude) signals 
in the low-frequency domain ($< 0.08$\,d$^{-1}$), (ii) merging and splitting signals over long timescales 
($0.08-0.15$\,d$^{-1}$), and (iii) signals with strongly variable amplitude with patchy, sometimes recurrent 
patterns ($> 0.15$\,d$^{-1}$). We averaged the frequencies over the time span and calculated their errors 
as the weighted standard deviation over a small window around each detected peak. These are listed in 
Table~\ref{tab:freq_mix} along with the frequencies obtained from the TESS light curves.
In this table, the frequencies are ordered with increasing values from top to bottom, and we aligned frequencies from different data categories that agree within their errorbars.
The same information is provided in graphical form in Fig.\,\ref{fig:freq-mix}, in which the symbol 
sizes additionally provide information about the power of the signals that were normalized to the 
highest power detected in the respective dataset, facilitating a direct comparison of relative signal strengths.

In addition, we tested whether pulsations are expected to be excited in HD~14134. For this, we used the GYRE 
stellar oscillation code \citep{2018MNRAS.475..879T} and computed non-adiabatic radial (l=0) and non-radial 
modes (l=1,2,3) within the range $0.01 < f~[\textrm{d}^{-1}] < 0.5$ for the two MESA models matching closely with the parameters 
of HD~14134 (see Table\,\ref{tab:mesa_models}). We found three excited non-radial modes in this frequency range for both M1 and M2. All predicted modes are included with their harmonic 
degree, $l$, in Table~\ref{tab:freq_mix}. 
The excited ones for $l=1$ in M1 have slightly lower frequencies but agree fairly well with those found for M2. 
The higher-frequency modes predicted by the individual models do not match. The one for M1  has no observational counterpart within the errorbars, whereas the one for M2 matches well with all the observations. Model M2 also has stellar parameters (mass, radius, luminosity) that fit closer to the observed ones (Table~\ref{tab:stelparam}), so that we consider the predictions from M2 to be representative for our object.

From model M2, we find one clear match for the theoretical frequency ($0.0521$\,d$^{-1}$, $P \simeq 19.2$\,d) with the 
values obtained from spectroscopy, which we identify as g-mode \citep{2006ApJ...650.1111S}. The mode is 
seen in the TESS data as well, but it aligns with our threshold value and is hence not listed in the table. 
Moreover, we find two more frequencies that match closely in the four data categories ($\sim0.1$\,d$^{-1}$; 
$\sim0.2$\,d$^{-1}$). They all have high powers (see Fig.\,\ref{fig:freq-mix}) but lack a theoretical 
counterpart. These frequencies seem to be multiples (harmonics) of the g-mode (2 and 4\,$f_{\rm g}$, but 
their frequencies and powers display modulations over the observing period). A third frequency is seen 
at least in the spectroscopic data ($\sim0.16$\,d$^{-1}$, although with lower powers) and might match 
with another harmonic (3\,$f_{\rm g}$). Inspection of the Fourier spectrum of the TESS data suggests that 
this signal is also present (see Fig.\,\ref{fig:enter-Fourier}), but its identification is suppressed by 
the high level of SLF variability in this frequency domain. 

Two frequencies in our table we consider as most likely due to wind variabilities as they are seen only in H$\alpha$ and TESS. These signals have periods of $\sim 7$\,d and $\sim 13$\,d, respectively. The latter
is similar to the $\sim 12.8$\,d signal reported by 
\citet{2004MNRAS.351..552M} based on their analysis of the \textsc{Hipparcos} photometry and equivalent 
width variations of the H$\alpha$ line. A third one with a period of $\sim 38$\,d is seen in H$\alpha$ and He\,{\sc i}, but not in  Si\,{\sc iii}. Therefore, we tend to assign it also more to the wind rather than to a pulsation mode, although similar periods are predicted by both M1 ($\sim 34$\,d) and M2 ($\sim 33$\,d).

We do not have an easy interpretation for the two high-frequency signals. These non-coherent signals appear in the scalograms as patchy or re-current signals with complex frequency structure and might be related to damped lifetimes of stochastic gravity waves \citep[e.g.][]{2019ApJ...876....4E}. We also have no explanation 
why none of the data sets shows indications for the predicted low-frequency pulsation modes. Although the 
scalograms of Si\,{\sc iii} and He\,{\sc i} seem to show some overlapping signals in this low-frequency 
domain, a clear separation into individual frequencies is difficult. Moreover, this frequency domain is the strongest impacted by the SLF variability. As we have seen, the properties of the SLF variability also vary with time,
suggesting that SLF variability is most likely connected to time-dependent processes in or close to the stellar surface 
layers. These could be caused by clumping and stellar wind inhomogeneities \citep{2011A&A...533A...4B, 2021MNRAS.502.5038N}, or by rotational modulation of gravity modes and the excitation of Rossby waves, as was recently suggested for a sample of late O-type supergiants \citep{Cidale2026}. 
However, a detailed investigation of the 
origin of the SLF variability phenomenon in HD~14134, and BSGs in general, requires more sophisticated analyses, which are beyond the scope of this work.

\subsection{Nature of HD~14134}

Our analysis of the spectra and SED of HD~14134 revealed that the values of its temperature, 
luminosity, radius, and current mass are in good agreement with the predictions from the MESA model for 
a star with an initial mass of 14\,M$_{\odot}$. As such, it is a post-main sequence object evolving 
towards the red-supergiant stage. This conclusion is in agreement with the age of the star that aligns 
with the one of its host cluster h~Per \citep{2021MNRAS.504..356D, 2026RAA....26d5016T}. Moreover, we do 
not detect radial pulsations in HD~14134, which is another argument for a pre-red supergiant evolutionary 
state of the star \citep{2013MNRAS.433.1246S}.

The derived stellar parameters result in a luminosity-to-mass ratio of
$(L/M) = 1.4\times 10^{4}\,(\textrm{L}_{\odot}/\textrm{M}_{\odot})$, which is in principle high enough 
for strange-mode instabilities to occur \citep[e.g.][]{2013MNRAS.433.1246S, 2024MNRAS.529.4947G}, and it 
seems plausible that some single-event or recurrent signals could be related to these instabilities. 
Strange modes were proposed to trigger phases of enhanced mass loss \citep{2016MNRAS.457.4330Y}, 
which is certainly a suitable mechanism to explain the wind variability seen in HD~14134, in particular 
the strong activity that we recorded at the beginning of the spectroscopic monitoring.

\section{Conclusions}
\label{sect:conclusions}

We presented an analysis of spectroscopic time-series and photometric light curves from the TESS mission for the BSG star HD~14134 with the aim to unveil the nature of the star and the physical cause for the observed variabilities. Our analysis revealed that HD~14134 is a post-main sequence object evolving towards the red-supergiant stage.

We found indication for stellar pulsations as well as wind-induced variability. In particular, we consistently detected a g-mode with a period of $\sim 19.2$\,d and its harmonics in all data sets. The mode was also predicted from theoretical models (using MESA and GYRE) for the stellar parameters that we have derived from the spectral and SED modelling.
The presence of g-mode pulsations, including their harmonics, was found in several BSGs \citep[e.g.][]{2006ApJ...650.1111S, 2007A&A...463.1093L, 2017A&A...602A..32A, 2019NatAs...3..760B}, supporting our identification.

The time-variable wind of the star is traced by the strong 
H$\alpha$ line variability. This wind variability, when transformed to photometric signals, can reach amplitudes on the same order as those observed by TESS. This finding emphasizes that variable stellar winds can have non-negligible contributions to light curves taken in broad-band (or white-light) filters that cover the wavelength region of the H$\alpha$ line.

Despite significant progress based on the excellent data from the TESS mission, disentangling between stellar pulsations and variability resulting from time-variable winds in BSGs still remains a
challenge. A considerable hindrance in our analysis were the single-sector TESS observations available for HD~14134, which do not allow to derive all frequencies with the needed precision. Longer cadence photometric observations, ideally paired with simultaneous spectroscopic monitoring, are clearly needed.
The upcoming PLATO space mission \citep{2025ExA....59...26R} that promises uninterrupted light curves and colour information is 
hence a very promising tool for future in-depth investigations.

\begin{acknowledgements}
We thank the reviewer for constructive comments on the manuscript.
This research made use of the NASA Astrophysics Data System (ADS) and of the SIMBAD database, 
operated at CDS, Strasbourg, France. This research has made use of the VizieR catalogue access 
tool, CDS, Strasbourg, France (DOI : 10.26093/cds/vizier). The original description 
of the VizieR service was published in 2000, A\&AS 143, 23.
Computational resources were provided by the e-INFRA CZ project (ID:90254), supported
by the Ministry of Education, Youth and Sports of the Czech Republic. Computational
resources were provided by the ELIXIR-CZ project (ID:90255), part of the international
ELIXIR infrastructure.
This project has received funding from the European Union (Project 101183150 - OCEANS) and from the Czech Science 
Foundation (GA \v{C}R, grant number 25-17532S).
The Astronomical Institute of the Czech Academy of Sciences is supported by the project 
RVO:67985815. 
\end{acknowledgements}

\bibliography{HD14134}
\bibliographystyle{aa}

\begin{appendix}
\nolinenumbers
\onecolumn

\section{Observation Log}

\begin{table*}[h!]
\centering   
\caption{Observation log of spectroscopic data. UT date is given in the format year-month-day. Heliocentric Julian Date (HJD) is computed in the middle of the (combined) exposures. S/N is measured in the vicinity of the H$\alpha$ line.}
\label{table:obs}
\begin{tabular}{ccc|ccc}  
\hline\hline
UT date  & HJD & S/N & UT date  & HJD & S/N \\ 
 & (day) &  &  & (day) &  \\
\hline											
2017-10-11	&	2458037.96725645	&	70	&	2017-12-04	&	2458091.8739076	&	84	\\
2017-10-12	&	2458038.96704434	&	80	&	2017-12-10	&	2458097.83683257	&	36	\\
2017-10-13	&	2458039.91190416	&	98	&	2017-12-12	&	2458099.89595775	&	60	\\
2017-10-14	&	2458040.91136581	&	102	&	2017-12-13	&	2458100.78801789	&	65	\\
2017-10-15	&	2458041.95062805	&	84	&	2017-12-15	&	2458102.78720172	&	83	\\
2017-10-16	&	2458042.90443349	&	97	&	2017-12-16	&	2458103.89687059	&	55	\\
2017-10-19	&	2458045.90338126	&	94	&	2017-12-20	&	2458107.89722266	&	66	\\
2017-10-23	&	2458049.90974458	&	96	&	2017-12-21	&	2458108.87702613	&	71	\\
2017-10-24	&	2458050.90904115	&	77	&	2017-12-22	&	2458109.88607719	&	44	\\
2017-10-25	&	2458051.90837908	&	87	&	2017-12-24	&	2458111.88316114	&	53	\\
2017-10-26	&	2458052.95178009	&	48	&	2017-12-25	&	2458112.88285759	&	78	\\
2017-10-27	&	2458053.82203571	&	93	&	2017-12-26	&	2458113.89954379	&	43\\	
2017-10-28	&	2458054.85820033	&	101	&	2017-12-28	&	2458115.88628559	&	67	\\
2017-10-29	&	2458055.91984851	&	73	&	2017-12-29	&	2458116.84177215	&	58	\\
2017-10-30	&	2458056.93461986	&	79	&	2017-12-30	&	2458117.87768132	&	59	\\
2017-11-03	&	2458060.91735185	&	71	&	2017-12-31	&	2458118.85603195	&	54	\\
2017-11-05	&	2458062.91574058	&	77	&	2018-01-02	&	2458120.86309253	&	35	\\
2017-11-06	&	2458063.67159516	&	96	&	2018-01-03	&	2458121.82700827	&	51	\\
2017-11-10	&	2458067.91389092	&	78	&	2018-01-05	&	2458123.83509441	&	57	\\
2017-11-11	&	2458068.91368308	&	66	&	2018-01-06	&	2458124.82401768	&	67\\	
2017-11-12	&	2458069.87045822	&	69	&	2018-01-07	&	2458126.48530104	&	108	\\
2017-11-13	&	2458070.69371944	&	80	&	2018-01-08	&	2458126.74506791	&	105	\\
2017-11-14	&	2458071.91204849	&	62	&	2018-02-06	&	2458155.6321693	&	65	\\
2017-11-15	&	2458072.91602665	&	62	&	2018-02-07	&	2458156.63251755	&	61	\\
2017-11-16	&	2458073.95748005	&	82	&	2018-02-08	&	2458157.63297034	&	110	\\
2017-11-17	&	2458074.95746185	&	83	&	2018-02-10	&	2458159.63416038	&	57	\\
2017-11-18	&	2458075.95746916	&	75	&	2018-02-11	&	2458160.63759349	&	67	\\
2017-11-19	&	2458076.95745086	&	67	&	2018-02-12	&	2458161.63474214	&	78	\\
2017-11-20	&	2458077.95744755	&	46	&	2018-02-13	&	2458162.60357523	&	75	\\
2017-11-21	&	2458078.95738346	&	60	&	2018-02-20	&	2458170.27277902	&	102	\\
2017-11-22	&	2458079.98077465	&	50	&	2018-03-02	&	2458179.61132823	&	77	\\
2017-11-23	&	2458080.98041942	&	50	&	2018-03-03	&	2458180.61180423	&	 70\\	
2017-11-26	&	2458084.00146832	&	49	&	2018-03-04	&	2458181.61212748	&	72	\\
2017-11-27	&	2458084.99761594	&	70	&	2018-03-05	&	2458182.61249374	&	73	\\
2017-11-28	&	2458085.99209371	&	30	&	2018-03-07	&	2458184.61347128	&	70	\\
2017-11-29	&	2458086.97815903	&	44	&	2018-03-12	&	2458189.61559439	&	82	\\
2017-12-03	&	2458090.87387006	&	78	&	\\					
\hline
\end{tabular}
\end{table*}

\pagebreak

\twocolumn

\section{Photometric light curves}

\begin{figure}[h!]
	\includegraphics[width=0.9\columnwidth]{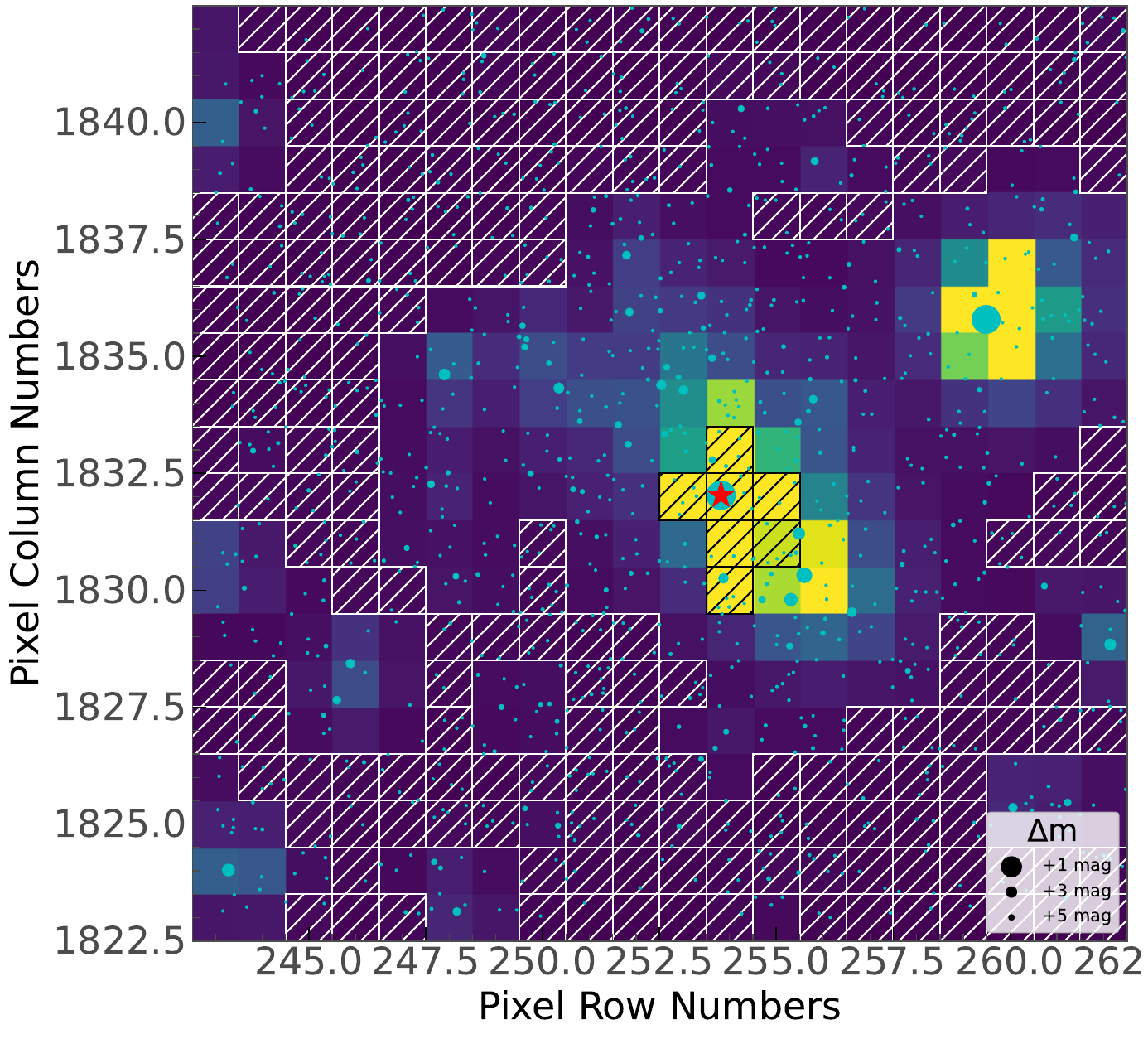}
    \includegraphics[width=0.9\columnwidth]{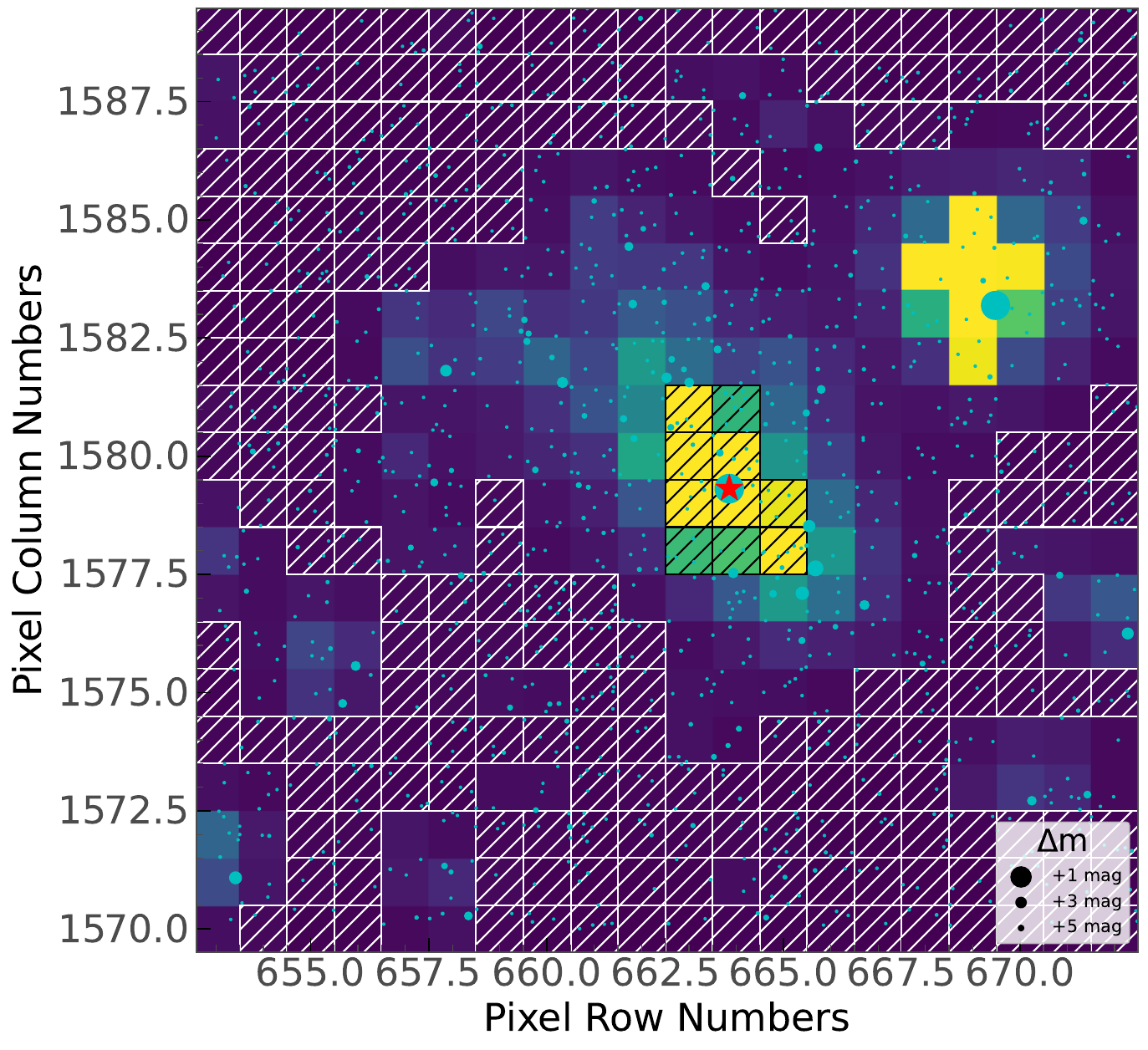}
    \caption{FFI cutouts of TESS sector 18 (\textit{top}) and 58 (\textit{bottom}). Shown are the positions of HD~14134 (\textit{red star}) and the background stars from Gaia DR2 (\textit{cyan dots}). Hatched boxes mark the target masking (\textit{black}) and the background masking (\textit{white}). The size of 
the dots indicates the magnitude difference $\Delta m$ of the Gaia sources with our star.}
    \label{fig:TPF_sectors}
\end{figure}

\begin{figure}[h!]
    \centering
	\includegraphics[width=0.9\columnwidth]{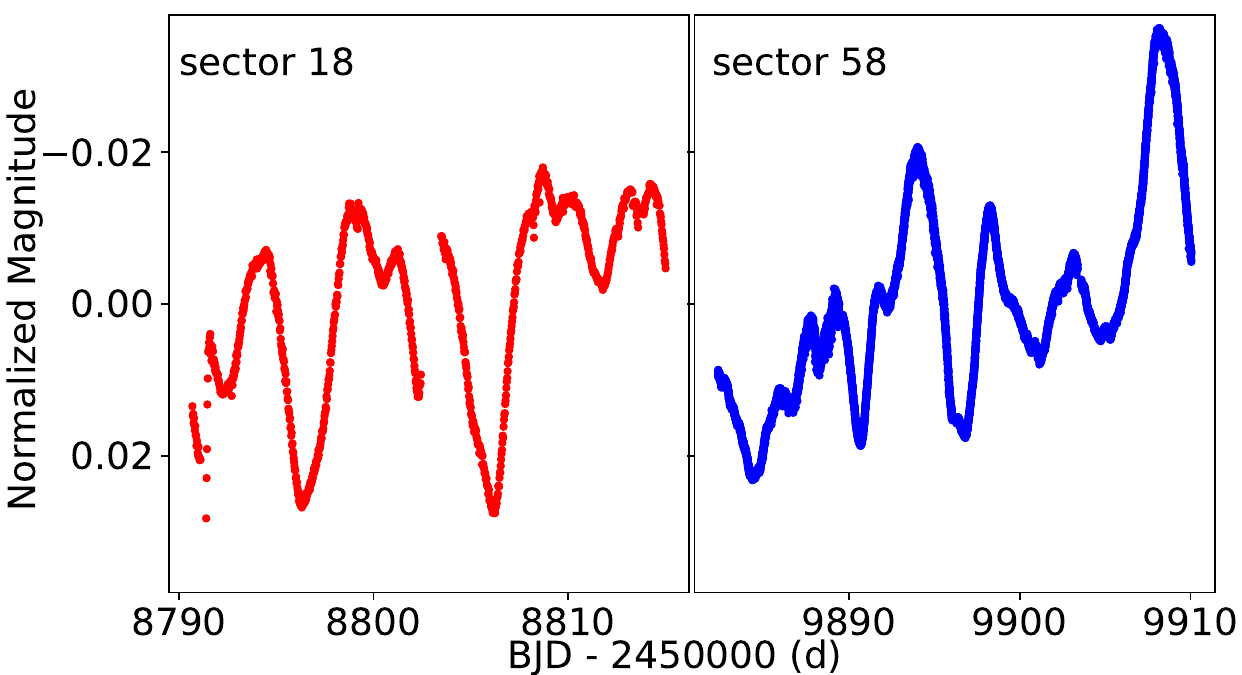}
    \caption{Extracted, normalized TESS light curves of HD~14134 from sector 18 (\textit{left}) and 58 (\textit{right}).}
     \label{fig:lc}
\end{figure}

\begin{figure}[h!]
    \centering
    \includegraphics[width=0.9\columnwidth]{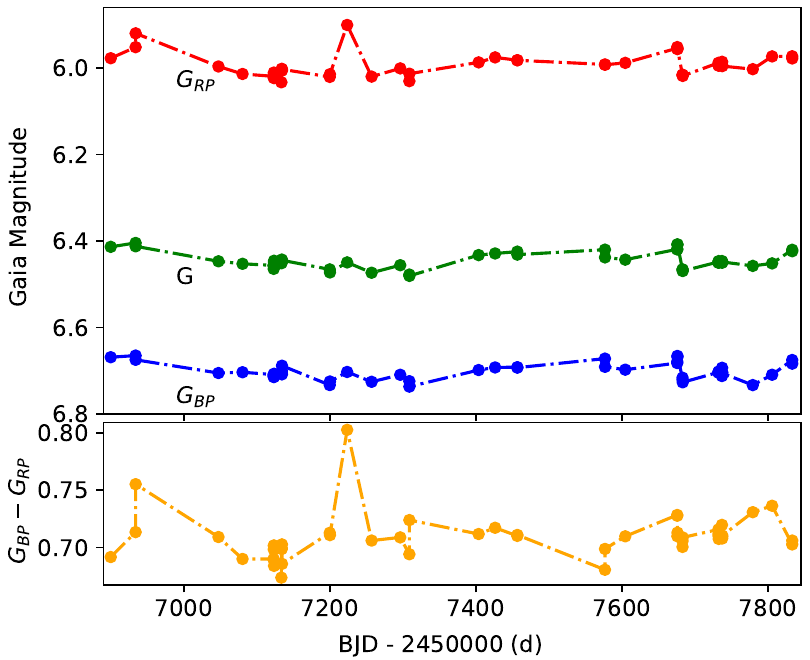}
    \caption{Light curves from Gaia DR3 in the three different bands (\textit{top}) and temporal colour variation (\textit{bottom}).}
    \label{fig:gaia}
\end{figure}

\begin{table}
\centering    
\caption{List of identified frequencies, their amplitudes, uncertainties, and S/N. Frequencies in brackets do not meet our S/N criterion. See text for more details on these.}             
\label{tab:FFT}
\begin{tabular}{cccccc}      
\hline\hline  
ID & Frequency & $3\sigma_{f}$ & Amplitude & $3\sigma_{A}$ & S/N  \\
& [d$^{-1}$] & [d$^{-1}$] & [mmag] & [mmag] & \\
\hline   
\multicolumn{6}{c}{Sector 18}    \\ 
   \hline            
   F1 & 0.2042 & 0.0017 & 12.311 & 0.781 & 8.50 \\ 
   F2 & 0.0927 & 0.0024 & 10.155 & 0.845 & 6.47 \\
  (F3) & 0.3129 & 0.0051 & 3.757 & 0.822 & 2.89 \\
  \hline
  \multicolumn{6}{c}{Sector 58} \\
\hline 
   (F4) & 0.1443 & 0.0013 & 9.175 & 0.002  &  4.28\\ 
   (F1) & 0.2000 & 0.0012 & 8.143 & 0.421 & 4.07\\
   \hline  
 \end{tabular}
\end{table}

\begin{table}
    \centering
    \caption{Frequencies obtained from the WWZ analysis. }
    \begin{tabular}{ccc}
    \hline \hline
   ID & Frequency (Sector 18) & Frequency (Sector 58)  \\
    & [d$^{-1}$] &  [d$^{-1}$] \\
    \hline
   f1 & 0.2050  & 0.2030  \\
   f2 & 0.0875   & -\\
   f3 & 0.3119   & - \\	
   f4 & 0.1296  & 0.1441  \\
   f5 & -  & 0.1059 \\     
    \hline
    \end{tabular}   
    \label{tab:wwz_freq}
\end{table}

\onecolumn

\section{Equivalent width measurements}

\begin{table*}[h!]
    \centering
    \caption{Equivalent width measurements of selected lines.}
    \label{tab:Eq_width}
    \begin{tabular}{lcccccccc}
    \hline
    \hline
     UT data & EW (Si\,{\sc ii}) & EW (Si\,{\sc ii}) & EW (Si\,{\sc iii}) & EW (Si\,{\sc iii}) & EW (He\,{\sc i}) & EW (He\,{\sc i}) & EW (Mg\,{\sc ii}) & EW (N\,{\sc ii})  \\
     (yyyy-mm-dd)  & (4128\,\AA) & (6347\,\AA) & (4552\,\AA) & (4568\,\AA) & (4471\,\AA) & (6678\,\AA) & (4481\,\AA) & (4601\,\AA) \\
     \hline
 2017-10-12 & 0.114 & 0.200 & 0.207 & 0.152 & 0.664  & 0.872  & 0.301  & 0.071\\
 2017-10-13 & 0.123 & 0.217 & 0.196 & 0.160 & 0.682  & 0.876  & 0.380  & 0.064\\
 2017-10-14 & 0.136 & 0.265 & 0.206 & 0.172 & 0.699  & 0.855  & 0.374  & 0.061\\
 2017-10-15 & 0.161 & 0.235 & 0.201 & 0.150 & 0.732  & 0.894  & 0.354  & 0.074\\
 2017-10-16 & 0.139 & 0.240 & 0.189 & 0.163 & 0.705  & 0.890  & 0.416  & 0.066\\
 2017-10-19 & 0.153 & 0.288 & 0.185 & 0.160 & 0.732  & 0.950  & 0.384  & 0.063\\
 2017-10-23 & 0.120 & 0.211 & 0.203 & 0.197 & 0.758  & 1.011  & 0.363  & 0.061\\
 2017-10-24 & 0.135 & 0.243 & 0.197 & 0.165 & 0.747  & 0.984  & 0.362  & 0.070\\
 2017-10-25 & 0.102 & 0.242 & 0.208 & 0.164 & 0.732  & 0.941  & 0.352  & 0.080\\
 2017-10-27 & 0.138 & 0.248 & 0.204 & 0.168 & 0.740  & 0.986  & 0.368  & 0.081\\
 2017-10-28 & 0.143 & 0.207 & 0.210 & 0.167 & 0.725  & 0.970  & 0.348  & 0.079\\
 2017-10-29 & 0.144 & 0.250 & 0.198 & 0.170 & 0.714  & 0.928  & 0.355  & 0.090\\
 2017-10-30 & 0.151 & 0.255 & 0.203 & 0.169 & 0.747  & 1.016  & 0.368  & 0.070\\
 2017-11-03 & 0.165 & 0.233 & 0.206 & 0.168 & 0.672  & 0.861  & 0.368  & 0.088\\
 2017-11-05 & 0.137 & 0.264 & 0.216 & 0.167 & 0.762  & 0.987  & 0.372  & 0.083\\
 2017-11-06 & 0.129 & 0.228 & 0.195 & 0.171 & 0.757  & 1.046  & 0.349  & 0.071\\
 2017-11-10 & 0.134 & 0.241 & 0.199 & 0.160 & 0.728  & 0.952  & 0.375  & 0.084\\
 2017-11-11 & 0.155 & 0.227 & 0.202 & 0.170 & 0.745  & 0.992  & 0.371  & 0.084\\
 2017-11-12 & 0.127 & 0.228 & 0.205 & 0.173 & 0.726  & 0.931  & 0.356  & 0.073\\
 2017-11-13 & 0.124 & 0.265 & 0.204 & 0.161 & 0.736  & 0.952  & 0.355  & 0.081\\
 2017-11-14 & 0.128 & 0.233 & 0.203 & 0.163 & 0.744  & 0.947  & 0.380  & 0.083\\
 2017-11-15 & 0.141 & 0.250 & 0.202 & 0.160 & 0.742  & 0.971  & 0.359  & 0.086\\
 2017-11-16 & 0.147 & 0.265 & 0.203 & 0.163 & 0.726  & 0.914  & 0.359  & 0.080\\
 2017-11-17 & 0.100 & 0.230 & 0.202 & 0.163 & 0.693  & 0.914  & 0.330  & 0.074\\
 2017-11-18 & 0.121 & 0.227 & 0.198 & 0.173 & 0.670  & 0.891  & 0.332  & 0.080\\
 2017-11-19 & 0.135 & 0.218 & 0.208 & 0.159 & 0.730  & 0.975  & 0.351  & 0.080\\
 2017-11-21 & 0.132 & 0.263 & 0.202 & 0.170 & 0.717  & 0.950  & 0.362  & 0.084\\
 2017-12-03 & 0.113 & 0.257 & 0.213 & 0.160 & 0.713  & 0.924  & 0.351  & 0.077\\
 2017-12-04 & 0.168 & 0.237 & 0.200 & 0.171 & 0.737  & 0.991  & 0.369  & 0.081\\
 2017-12-12 & 0.099 & 0.243 & 0.201 & 0.160 & 0.756  & 0.997  & 0.385  & 0.083\\
 2017-12-13 & 0.122 & 0.242 & 0.196 & 0.153 & 0.681  & 0.878  & 0.336  & 0.086\\
 2017-12-15 & 0.133 & 0.246 & 0.197 & 0.156 & 0.736  & 0.905  & 0.359  & 0.067\\
 2017-12-20 & 0.136 & 0.283 & 0.206 & 0.170 & 0.734  & 0.995  & 0.367  & 0.074\\
 2017-12-21 & 0.099 & 0.270 & 0.195 & 0.159 & 0.706  & 0.930  & 0.348  & 0.072\\
 2017-12-25 & 0.103 & 0.273 & 0.186 & 0.151 & 0.703  & 0.913  & 0.357  & 0.074\\
 2018-12-28 & 0.125 & 0.252 & 0.202 & 0.168 & 0.705  & 0.885  & 0.342  & 0.090\\
 2018-01-06 & 0.121 & 0.286 & 0.226 & 0.172 & 0.777  & 1.036  & 0.384  & 0.075\\
 2018-01-07 & --    & 0.222 & 0.198 & 0.147 & 0.734  & 0.944  & 0.345  & 0.062\\
 2018-01-08 & --    & 0.218 & 0.200 & 0.179 & 0.766  & 0.936  & 0.345  & 0.072\\
 2018-02-06 & 0.117 & 0.266 & 0.204 & 0.168 & 0.719  & 0.908  & 0.332  & 0.072\\
 2018-02-07 & 0.118 & 0.252 & 0.195 & 0.159 & 0.707  & 0.923  & 0.346  & 0.073\\
 2018-02-08 & --    & 0.221 & 0.202 & 0.169 & 0.708  & 0.875  & 0.334  & 0.078\\
 2018-02-11 & 0.116 & 0.260 & 0.207 & 0.154 & 0.693  & 0.880  & 0.338  & 0.069\\
 2018-02-12 & 0.132 & 0.249 & 0.195 & 0.158 & 0.696  & 0.873  & 0.357  & 0.070\\
 2018-02-13 & 0.102 & 0.239 & 0.208 & 0.163 & 0.718  & 0.926  & 0.364  & 0.079\\
 2018-02-20 & --    & 0.256 & 0.240 & 0.171 & 0.785  & 1.044  & 0.395  & 0.072\\
 2018-03-02 & 0.116 & 0.248 & 0.220 & 0.173 & 0.702  & 0.917  & 0.347  & 0.068\\
 2018-03-03 & 0.122 & 0.246 & 0.198 & 0.153 & 0.703  & 0.884  & 0.341  & 0.075\\
 2018-03-04 & 0.126 & 0.228 & 0.201 & 0.167 & 0.699  & 0.897  & 0.337  & 0.079\\
 2018-03-05 & 0.115 & 0.245 & 0.203 & 0.173 & 0.724  & 0.910  & 0.336  & 0.067\\
 2018-03-07 & 0.135 & 0.266 & 0.208 & 0.163 & 0.706  & 0.916  & 0.342  & 0.074\\
 2018-03-12 & 0.147 & 0.228 & 0.192 & 0.151 & 0.679  & 0.895  & 0.330  & 0.082\\ 
 \hline 
 <EW>       & 0.129 & 0.244 & 0.203 & 0.164 & 0.721  & 0.935  & 0.356  & 0.076 \\
 $\sigma$   & 0.017 & 0.020 & 0.009 & 0.009 & 0.027  & 0.049  & 0.019  & 0.007\\
 \hline
  \hline
 \end{tabular}
\end{table*}

\section{Line-profile variability}

\begin{figure*}[h!]
    \centering
    \includegraphics[width=0.88\textwidth]{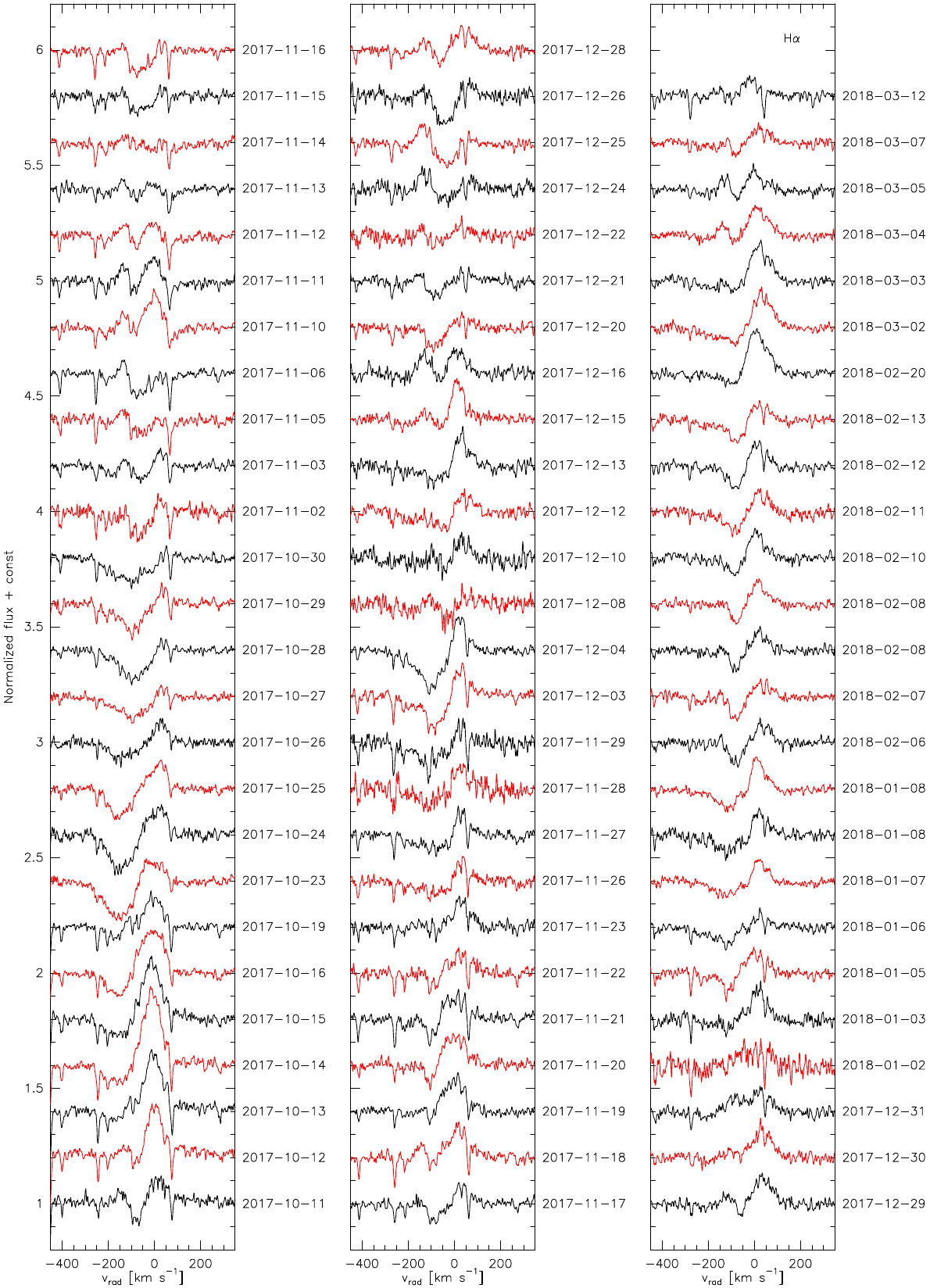}
      \caption{Temporal evolution of the H$\alpha$ profile.}
      \label{fig:Ha_profile}   
\end{figure*}

\end{appendix}

\end{document}